\documentclass{aa}  

\usepackage{graphicx}
\usepackage{txfonts}
\usepackage{lscape}
\usepackage{color}

\begin{document} 

  \title{The \textsc{Stagger}-grid: A grid of 3D stellar atmosphere models}
  \subtitle{V. Synthetic stellar spectra and broad-band photometry}
\titlerunning{The \textsc{Stagger}-grid -- V. Synthetic stellar spectra and broad-band photometry}

  \author{A. Chiavassa \inst{1}, L. Casagrande\inst{2}, R. Collet\inst{3}, Z. Magic\inst{4,5}, L. Bigot\inst{1}, F. Th\'evenin \inst{1}, M. Asplund\inst{2}}

\authorrunning{A. Chiavassa et al.}
\institute{Universit\'e C\^ote d'Azur, Observatoire de la C\^ote d'Azur, CNRS, Lagrange, CS 34229, Nice,  France \\
\email{andrea.chiavassa@oca.eu}
\and
Research School of Astronomy $\&$ Astrophysics, Australian National University, Cotter Road, Weston ACT 2611, Australia
\and
Stellar Astrophysics Centre, Department of Physics and Astronomy, Ny Munkegade 120,  Aarhus University, DK-8000 Aarhus C, Denmark
\and
Niels Bohr Institute, University of Copenhagen, Juliane Maries Vej 30, DK--2100 Copenhagen, Denmark  
\and 
Centre for Star and Planet Formation, Natural History Museum of Denmark, University of Copenhagen, {\O}ster Voldgade 5-7, DK--1350 Copenhagen, Denmark}
 
   \date{...; ...}

 
  \abstract
   {The surface structures and dynamics of cool stars are characterised  by the presence of convective motions and turbulent flows which shape the emergent spectrum.}
   {We used realistic three-dimensional (3D)  radiative hydrodynamical simulations from the \textsc{Stagger}-grid to calculate synthetic spectra with the radiative transfer code {{\sc Optim3D}} for stars with different stellar parameters to predict photometric colours and convective velocity shifts.}
   {We calculated spectra  from 1000 to 200\ 000 \AA\ with a constant resolving power of $\lambda/\Delta\lambda=$20\ 000 and   from 8470 and 8710 \AA\ (Gaia Radial Velocity Spectrometer - RVS - spectral range) with a constant resolving power of $\lambda/\Delta\lambda=$300\ 000.}
   {We used synthetic spectra to compute theoretical colours in the Johnson-Cousins $UBV(RI)_C$, SDSS, 2MASS, Gaia, SkyMapper, Str\"omgren systems, and HST-WFC3. Our synthetic magnitudes are compared with those obtained using  1D hydrostatic models. We showed that 1D versus 3D differences are limited to a small percent except for the narrow filters that span the optical and UV region of the spectrum. In addition, we derived the effect of the convective velocity fields on selected Fe $\mathrm{I}$ lines. We found the overall convective shift for 3D simulations with respect to the reference 1D hydrostatic models, revealing line shifts of between -0.235 and +0.361 km/s. We showed a net correlation of the convective shifts with the effective temperature: lower effective temperatures denote redshifts and higher effective temperatures denote blueshifts. We conclude that the extraction of accurate radial velocities from RVS spectra need an appropriate wavelength correction from convection shifts.}   
   {The use of realistic 3D hydrodynamical stellar atmosphere simulations has a small but significant impact on the predicted photometry compared with classical 1D hydrostatic models for late-type stars. We make all the spectra publicly available for the community through the POLLUX database.}
   

   \maketitle

%

\section{Introduction}

The stellar atmosphere is the boundary to the opaque stellar interior, and serves as the link between observations and the models of stellar structure and evolution. The phenomena of stellar evolution manifest themselves in the stellar surface as changes in chemical composition and in fundamental stellar parameters such as radius, surface gravity, effective temperature, and luminosity. The information we use to study distant stars comes from the flux they have emitted. However, the atmospheric layers where this flux forms is the transition region between convective and radiative regime. Thus, the surface structures and dynamics of cool stars are characterised by the presence of convective motions and turbulent flows. Convection manifests in the surface layers as a particular pattern of downflowing cooler plasma and bright areas where hot plasma rises \citep{2009LRSP....6....2N}. The size of granules depends on the stellar parameters of the star and, as a consequence, on the extension of their atmosphere  \citep[e.g. ][]{2013arXiv1302.2621M}. Eventually, the convection causes an inhomogeneous stellar surface that changes with time. They affect the atmospheric stratification in the region where the flux forms and also affect the emergent spectral energy distribution (SED), with potential effects on the precise determinations of stellar parameters \citep[e.g. ][]{2011A&A...534L...3B,2012A&A...545A..17C, 2012A&A...540A...5C}, radial velocity \citep[e.g. ][]{2008sf2a.conf....3B,2011JPhCS.328a2012C,2013A&A...550A.103A}, chemical abundance \citep[e.g. ][]{2005ASPC..336...25A,2009ARA&A..47..481A,2011SoPh..268..255C}, photometric colours \citep{2017MmSAI..88...90B}, and on planet detection \citep{2015A&A...573A..90M,2017A&A...597A..94C}. 

Convection is a difficult process to understand because it is  non-local, and 3D, and it involves non-linear interactions over many disparate length scales. In this context, the use of numerical  3D radiative hydrodynamical simulations of stellar convection is extremely important. In recent years, with increased computational power, it has been possible to compute grids of 3D simulations that cover a substantial portion of the Hertzsprung--Russell diagram \citep{2013arXiv1302.2621M,2013ApJ...769...18T,2009MmSAI..80..711L}. With these tools it is possible to predict reliable synthetic spectra for several stellar types.

Photometric systems and filters are designed to be sensitive to temperature, gravity, and metal abundance indicators and thereby to complement spectroscopic determinations of the fundamental properties of stars. In addition, the integrated magnitudes and colours of stars can be used to infer the ages, metallicities, and other properties of the underlying stellar populations \citep[e.g. ][]{2014MNRAS.444..392C}. For these purposes, several broad-band, or intermediate- and/or narrow-band filters have been designed to probe different regions of stellar spectra sensitive to different atmospheric parameters \citep{2012PASP..124..140B,2006AJ....131.2332G,2003AJ....126.1090C}. Additionally, there are the photometric systems used by the Gaia mission. \\
Gaia \citep{2016A&A...595A...1G} is an astrometric, photometric, and spectroscopic spaceborne mission of a large part of the Milky Way. Apart from the astrometric instrument, Gaia carries on board two low-resolution spectrophotometers \citep[Blue and Red Prism, BP/RP, ][]{2013A&A...559A..74B} and the Radial Velocity Spectrometer \citep[RVS, ][]{2004MNRAS.354.1223K}. The photometric instrument measures the SED over 120 pixels of all detected objects at the same angular resolution and at the same epoch as the astrometric observations. The BP operates in the range 3300-6800 \AA, while the RP uses the range 6400-10500 \AA \ . The main aims of this instrument are to provide  proper classifications (e.g. distinguish between stars, galaxies, and quasars) and characterisations (e.g. reddenings and stellar parameters), and  to enable chromatic corrections of the astrometric centroid data. Finally, the integral-field spectrograph RVS provides spectra between 8470 and 8710 \AA\ at a spectral resolving power of $\approx$ 11\ 200. The RVS is expected to produce  radial velocities through Doppler-shift measurements;  interstellar reddening, atmospheric parameters, and projected rotational velocities; and  individual element abundances for some elements. 

In this work, we calculated synthetic stellar spectra and photometry for broad-band and Gaia systems obtained using realistic 3D radiative hydrodynamical simulations of stellar convection from the  \textsc{Stagger}-grid. The predicted photometric 3D colours and the 3D spectra are publicly available for the community through the POLLUX database. The spectra corresponding to the RVS/Gaia spectral range will be used for the calibration of the instrument to preserve the measured radial velocities from the convection shift (see forthcoming paper of GaiaDataRelease2/RVS).
  
   \begin{figure*}
   \centering
    \begin{tabular}{c}
       \includegraphics[width=0.77\hsize]{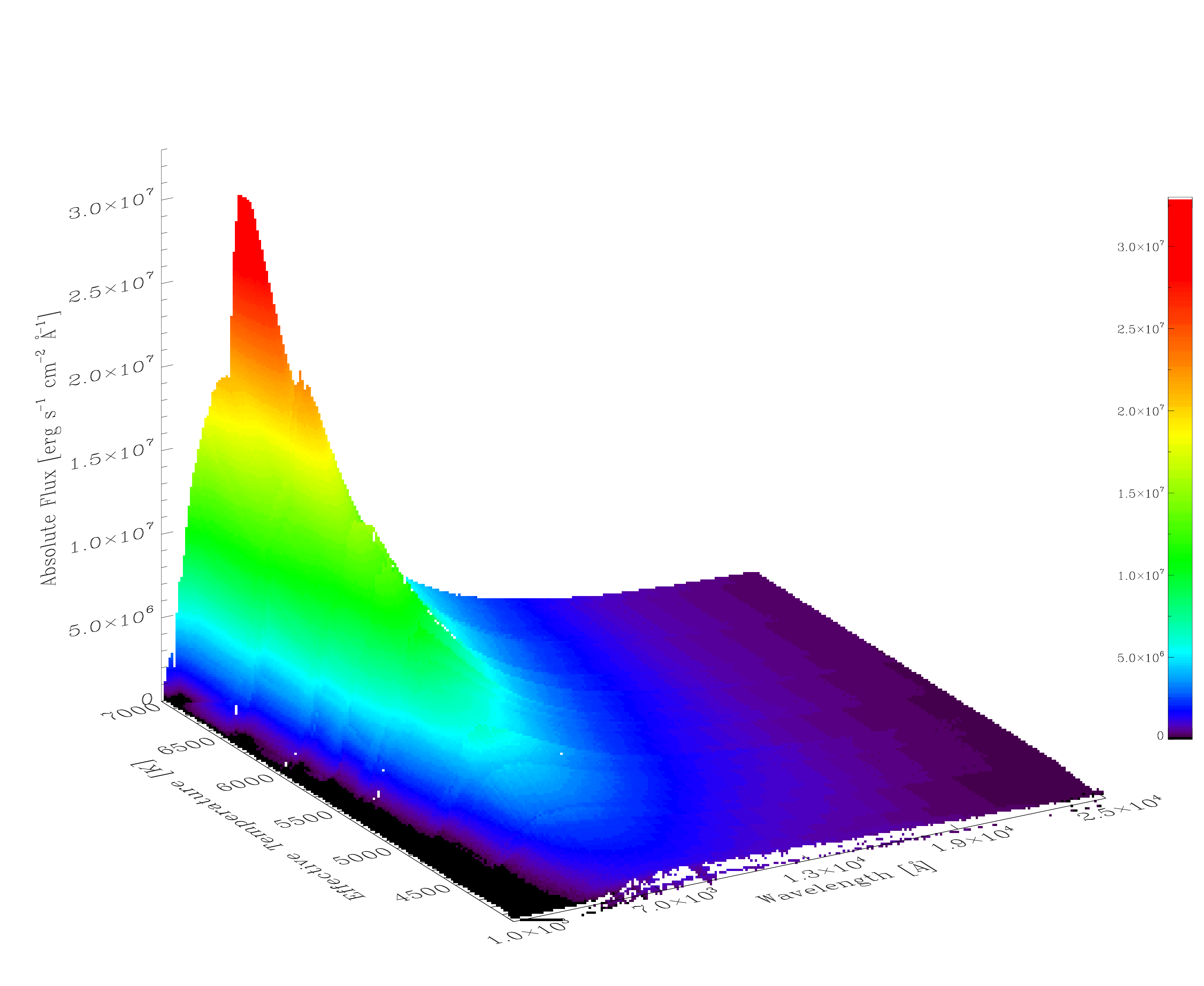} \\
  \end{tabular}
      \caption{Surface rendering for all the synthetic spectra computed for the 3D RHD simulations in Table~4. The vertical bar on the right displays the colour scale for the emerging flux in erg/s/cm$^2$/\AA. For clarity, the wavelength  range has been reduced to 1000 -- 25\ 000 \AA\
           }
        \label{first}
           \end{figure*}

\section{Methods}
      
   \subsection{Stellar model atmospheres}
   
  \cite{2013arXiv1302.2621M} described the \textsc{Stagger}-grid of realistic 3D radiative hydrodynamical simulations of stellar convection for cool stars using the \textsc{Stagger}-code (originally developed by Nordlund $\&$ Galsgaard 1995\footnote{http://www.astro.ku.dk/$\sim$kg/Papers/MHDcode.ps.gz}, and continuously improved over the years by its user community), a state-of-the-art (magneto)hydrodynamic code that solves the time-dependent hydrodynamic equations for mass-, momentum-, and energy-conservation, coupled with the 3D radiative transfer equation in order to account correctly for the interaction between the radiation field and the plasma. The code uses periodic boundary conditions horizontally and open boundaries vertically. At the bottom of the simulation, the inflows have a constant entropy. The outflows are not tightly constrained and are free to pass through the boundary. The code is based on a sixth-order explicit finite-difference scheme and a fifth-order interpolation. The considered large number over wavelength points is merged into 12 opacity bins \citep{1982A&A...107....1N,2000ApJ...536..465S,2013arXiv1302.2621M}. \textsc{Stagger} simulations are based on a realistic equation of state that accounts for ionisation, recombination, and dissociation \citep{MHD}; continuous absorption and scattering coefficients listed in \cite{2010A&A...517A..49H}; and the line opacities listed in \cite{2008A&A...486..951G}. These are in turn based on the VALD-2 database \citep{2001ASPC..223..878S} of atomic lines and the SCAN-base \citep{1997IAUS..178..441J} of molecular lines. 
   
    \subsection{Three-dimensional radiative transfer}
    
    We used the 3D pure-LTE radiative transfer code \textsc{Optim3D} \citep{2009A&A...506.1351C} to compute synthetic spectrum from the snapshots of the RHD simulations of the \textsc{Stagger}-grid
     \citep[see Fig.~1 in ][]{2013arXiv1302.2621M}. The code takes into account the Doppler shifts due to convective motions. The radiative transfer equation is solved monochromatically using pre-tabulated extinction coefficients as a function of temperature, density, and wavelength. \\
    The lookup tables were computed for the same chemical compositions as the RHD simulations using the same extensive atomic and molecular continuum and line opacity data as the latest generation of MARCS models \citep{2008A&A...486..951G} with the addition---with respect to Table 2 of Gustafsson's paper---of the SiS molecule \citep{2009ApJ...690L.122C}, which is particularly important for the far-infrared region of the spectrum. While the sources of line opacities used in the RHD simulations of \cite{2013arXiv1302.2621M} and in \textsc{Optim3D} are the same, the data for the continuum opacities are almost the same: \cite{2010A&A...517A..49H} reported that the data used in RHD simulations are mostly identical to those used in the MARCS models, but include additional bound-free data from the Opacity Project and the Iron Project (Trampedach et al., private communication) as well as some opacities of the second ionisation stage for many metals.\\
    For the computation of the spectra from RHD simulations, the assumed microturbulence is equal to zero since the velocity fields inherent in RHD models are expected to self-consistently and adequately account for non-thermal Doppler broadening of spectral lines \citep{2000A&A...359..755A}. The temperature and density ranges spanned by the tables are optimised for the values encountered in the RHD simulations. The detailed methods used in the code are explained in \cite{2009A&A...506.1351C,2010A&A...524A..93C}. \textsc{Optim3D} has already been employed in synergy with the \textsc{Stagger} simulations in several works \citep{2010A&A...524A..93C,2011JPhCS.328a2012C,2012A&A...540A...5C,2014A&A...567A.115C,2015A&A...573A..90M,2015A&A...576A..13C,2017A&A...597A..94C} either concerning the extraction of synthetic spectra or interferometric observables.
    
     \subsection{One-dimensional radiative transfer}
    
    For all the following comparisons with 3D simulations, we used plane-parallel, hydrostatic, 1D atmosphere models computed with a similar physical treatment to the MARCS code and the same equation of state  and opacities as in the individual 3D simulations \citep[ATMO, ][]{2013arXiv1302.2621M}. Moreover, we used a 1D version of \textsc{Optim3D,} and the chemical compositions, the opacities, and numerics of the radiative transfer calculations for the emergent intensities are the same as used in the 1D and 3D approaches. 
        
   \section{Synthetic spectra from 0.1 to 20 $\mu$m}
 
   \begin{figure}
   \centering
    \begin{tabular}{cc}
    \includegraphics[width=0.9\hsize]{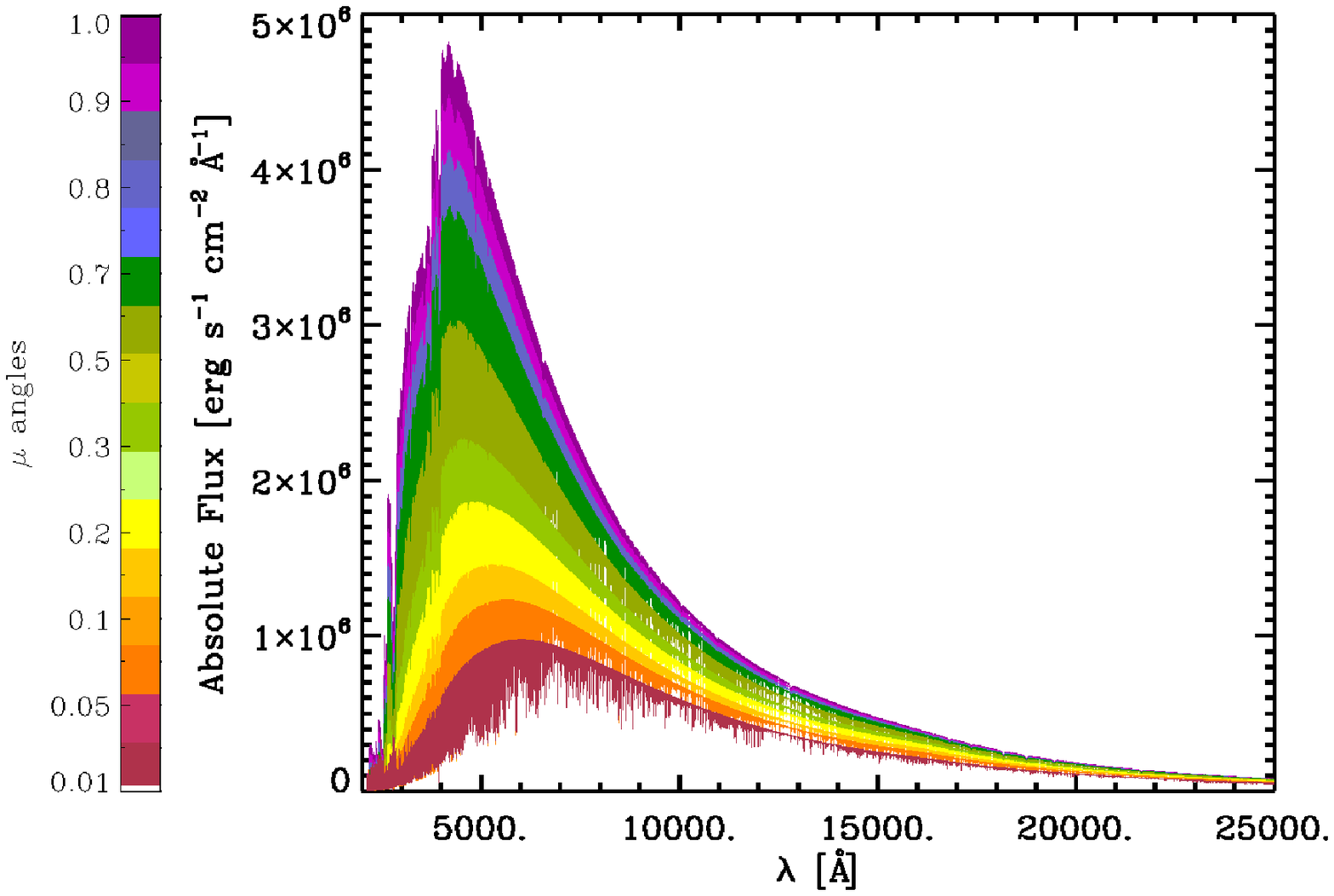} 
  \end{tabular}
      \caption{Synthetic spectra of the solar simulation in the spectral range 2000 -- 25\ 000 \AA\ and for the different $\mu=\cos(\theta)$ inclination angles used in the computation, where $\theta$ is the angle with respect to the line of sight (vertical axis).}
        \label{spectraexample}
           \end{figure}
   

The \textsc{Stagger}-grid includes 3D stellar atmosphere simulations with metallicities [Fe/H]  = $+$0.5, 0.0, $-$0.5, $-$1.0, $-$2.0, $-$3.0, and $-$4.0; surface gravity log$g$ between 1.5 and 5.0 in steps of 0.5 dex; and effective temperature $T_{\rm{eff}}$ from 4000 to 7000 K in steps of 500 K (Fig.~1 of Paper I). In this work we present the synthetic spectra computed for the \textsc{Stagger}-grid for a total of 181 simulations (Table~4). The spectra have been calculated with a constant resolving power of $\lambda/\Delta\lambda=$20\ 000 ($n_\lambda$ = 105\ 767 wavelength points) from 1000 to 200\ 000 \AA. \textsc{Optim3D} computes the emerging intensities for vertical rays cast through the computational box for all required wavelengths. The procedure is repeated after tilting the computational box by an angle $\theta$ with respect to the line of sight (vertical axis) and rotating it azimuthally by an angle $\phi$. The final result is a spatially resolved intensity spectrum at different angles. In addition,  a temporal average is also performed. We performed the calculations for ten snapshots of the 3D RHD simulations of Table~4, adequately spaced  so as to capture several convective turnovers, for ten different inclination angles $\mu=\cos(\theta)$ = $[1.00, 0.90, 0.80, 0.70, 0.50, 0.30, 0.20, 0.10, 0.05, 0.01]$ (see Fig.~\ref{spectraexample}), and four $\phi$-angles [0$^\circ$, 90$^\circ$, 180$^\circ$, 270$^\circ$]. The strongest decline in the limb darkening is usually found towards the limb; therefore, we decided to resolve with more $\mu$-angles at the limb instead of having an equidistant scale in $\mu$.  We tested the discrepancy between the temporal average using a large number of snapshots (e.g. 20) and using only 10 snapshots is lower than 0.3$\%$. The number of ten snapshots was chosen because it represents the best compromise in terms of computational time and accuracy among the whole set of stellar parameters. All things considered, we computed 400 spectra in the range 1000 -- 200\ 000 \AA\ for every simulation.

       \begin{figure}
   \centering
    \begin{tabular}{c}
   \includegraphics[width=0.98\hsize]{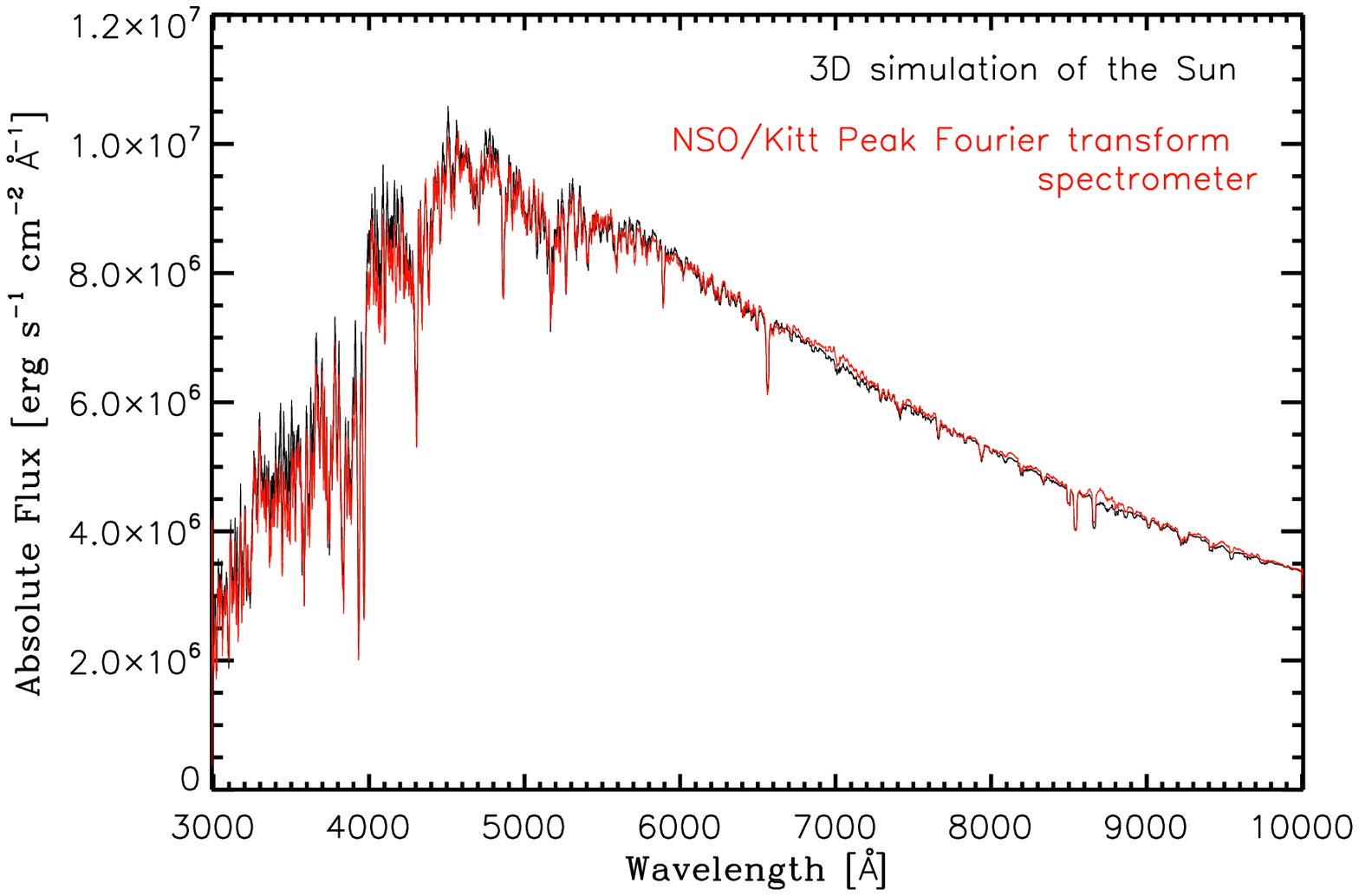} 
  \end{tabular}
      \caption{Comparison of the solar simulation (black) with the observed flux of the Sun \citep[red, ][]{2005MSAIS...8..189K}. The solar irradiance is converted to flux at the solar surface using the multiplicative factor of $\left[\left(1 \rm{AU}\right)/\rm{R}_{\odot}\right]^2 = 46202$. For clarity, the spectra have been resampled to a lower spectral resolution with fewer frequency points ($n_\lambda$ = 2115).}
        \label{allspectra2}
           \end{figure}

Figure~\ref{first} displays the set of all synthetic spectra computed.  We determined the $T_{\rm{eff}}$ from the integration of the SED of the spectra from 0.1 to 20 $\mu$m. The effective temperature has been computed using Stefan--Boltzmann law as
\begin{equation}
 T_{\rm{eff}\rm{\_spectra}} = \left\{\left[\int_{\lambda_1}^{\lambda_2}f\left(\lambda\right)d\lambda\right]/\sigma\right\}^{0.25}
,\end{equation}
 where $\lambda_1=$1010\AA\ and $\lambda_2$=199\ 960 \AA\ , $f\left(\lambda\right)$ is the synthetic flux, and $\sigma$ is the Stefan--Boltzmann constant. The values of the effective temperature are listed in Table~4. \\
 Figure~\ref{first} shows that increasing $T_{\rm{eff}}$ returns higher radiated energy per surface area and the peak of the radiation curve moves to shorter wavelengths, as expected by Planck law. \cite{2013A&A...554A.118P} provided excellent agreement of their 3D solar simulation of the \textsc{Stagger}-grid with the continuum observation of the Sun. As they did, we used the \cite{2005MSAIS...8..189K} irradiance\footnote{Available at http://kurucz.harvard.edu/sun.html} and normalised flux atlases for the Sun between 3000 and 10000 \AA\ and found a good agreement (Fig.~\ref{allspectra2}), reinforcing the view that the simulation of thermodynamic structure and post-processing detailed radiative transfer are realistic. This conclusion was also reported by \cite{2012A&A...539A.102H}, who determined that the numerical resolution of the \textsc{Stagger} 3D RHD models and the spectral resolution for the flux computations are sufficient to predict realistic observables. In particular, some of the RHD simulations presented in this work and for a limited spectral region between 2000 and 10\ 000 \AA\ have been used in \cite{2015A&A...573A..90M} to provide appropriate coefficients for various bi-parametric and non-linear limb darkening laws.

 \section{Photometric synthetic observables}

Photometric systems and filters are designed to probe fundamental physical parameters, such as the effective temperature, surface gravity, and metallicity of stars. Colour and magnitude relations are used for a variety of purposes:  interpreting the observed distribution of stars in colour-colour and colour-magnitude diagrams,  deriving distances to stars and star clusters, and testing stellar evolutionary theory by comparing with observations  to name just a few. Thus, it is important to have realistic model fluxes to generate colours which match the observed values. In addition to synthetic model fluxes,  details on the photometric standardisation are also a part of  this quest. 

In essence, photometry condenses the information encoded in a spectrum
$f\left(\lambda\right)$ over a system response function $T\left(\lambda\right)$,
 i.e. $\int f\left(\lambda\right) T\left(\lambda\right) d\lambda$. Each existing
photometric system then varies in the details. Most notably,
$T\left(\lambda\right)$ will depend on the filter under consideration and the
response function of the detector. This means that  a distinction must be made
between photo-counting and energy-integration detectors, meaning that a
measurement of energy
$\int f\left(\lambda\right) T\left(\lambda\right) d\lambda$ will correspond to
$(hc)^{-1} \int f\left(\lambda\right) \lambda T\left(\lambda\right) d\lambda$
photons \citep[see e.g. ][]{2000PASP..112..961B}
Another aspect that often varies among different photometric systems is how their
standardisation (zero-point and absolute calibration) is achieved. Here, for
all systems but Gaia we adopted the exact same procedure as used by \cite{2014MNRAS.444..392C}, where details on the adopted filter transmission curves, the
photo-counting and energy-integration formalism, and zero-points and absolute
calibration can be found\footnote{The only difference with respect to
  \cite{2014MNRAS.444..392C} is that here we have adopted
  $M_{\rm{Bol}\odot}=4.74$}. We computed synthetic colours in the
Johnson-Cousins, SDSS, 2MASS, Gaia, SkyMapper, Str\"omgren, HST-WFC3, and Gaia
systems (Table~\ref{tablefilters} and Fig.~\ref{spectraexample} for a comparison of the solar spectrum with the filter
transmission curves studied here). For the HST-WFC3 systems our tables are
provided in the VEGA, ST, and AB systems.

A full characterisation of the Gaia photometric system, including zero-points
and standardisation is expected to be released in 2018. In this work, we used
the transmission curves available from the ESA website\footnote{https://www.cosmos.esa.int/web/gaia/transmissionwithoriginal}, and computed Gaia colours following \cite{2010A&A...523A..48J}. We fixed Vega's magnitudes to be G=BP=RP=0.03, and for the absolute calibration used a Kurucz synthetic Vega spectrum rescaled to the measured flux value at 5556 \AA\ from \cite{1995A&A...296..771M}.

Similarly to \cite{2014MNRAS.444..392C}, instead of colour indices we
provide bolometric corrections in different bands (Table 4 -- Table 8) because they
are more versatile and can  be rearranged in any colour combination, as follows
from Eqs.~(\ref{eq:bc1}) and~(\ref{eq:bc2}). The bolometric magnitude is defined
as
\begin{equation}
M_{\rm{Bol}} = -2.5 \log\frac{L}{L_\odot} + M_{\rm{Bol},\odot},
\end{equation}
where we adopt $M_{\rm{Bol},\odot}$ = 4.74. It follows that the bolometric correction in a given band $BC_\zeta$ is
\begin{equation}\label{eq:bc1}
BC_\zeta = m_{Bol}-m_\zeta=M_{Bol}-M_\zeta,
\end{equation}
where the lower and upper cases refer to apparent and absolute magnitudes, respectively. From this it follows that colour indices can be obtained from the difference in bolometric corrections, where $\zeta$ and $\eta$ are two given bands:
\begin{equation}\label{eq:bc2}
\zeta - \eta = m_\zeta - m_\eta = BC_\eta - BC_\zeta.
\end{equation}
Thus,  in the rest of the paper when we talk about synthetic colours, these
have been obtained as differences in bolometric corrections from our tables.

    \begin{figure*}[!h]
   \centering
    \begin{tabular}{c}
    \includegraphics[width=0.47\hsize]{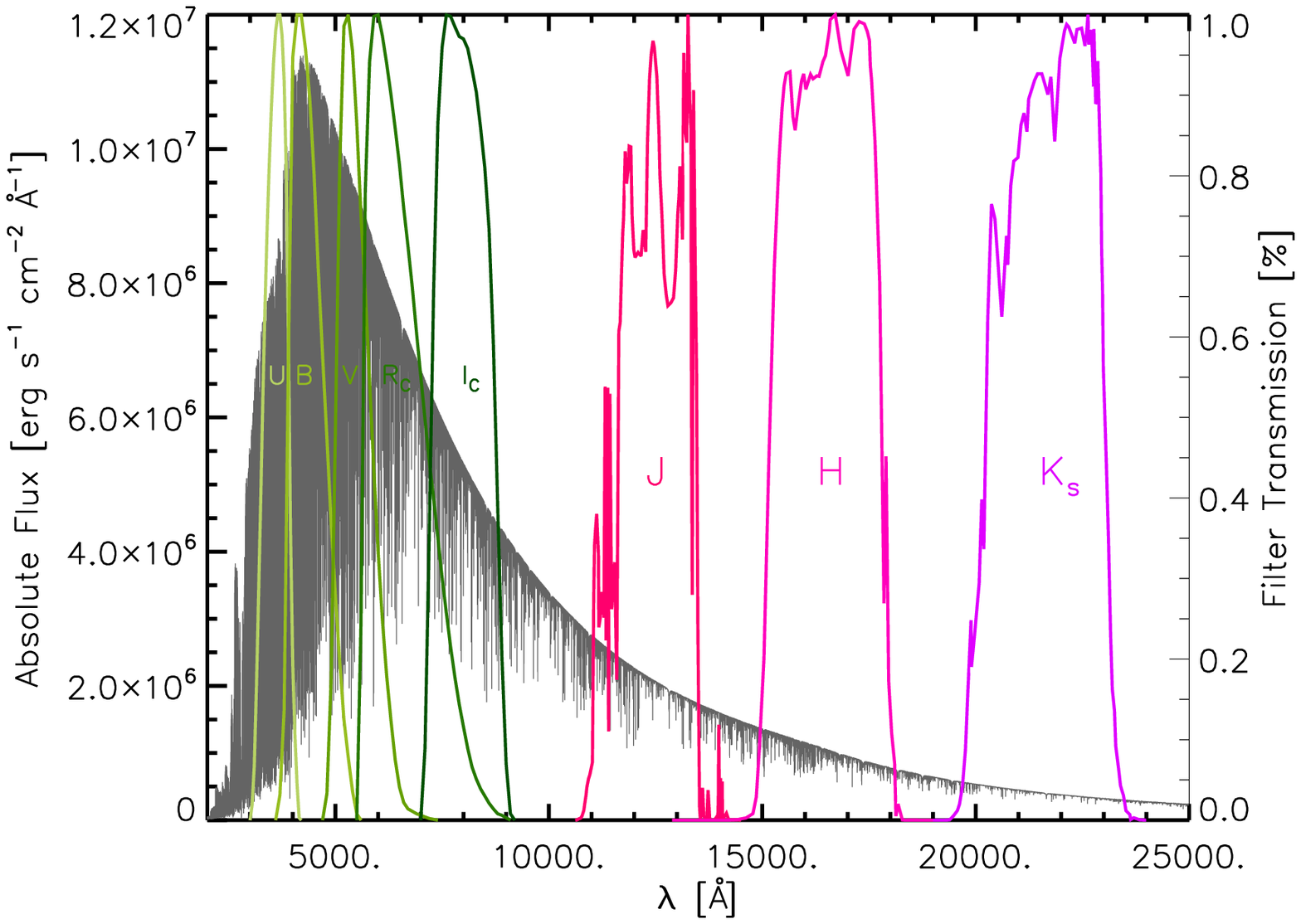} 
     \includegraphics[width=0.47\hsize]{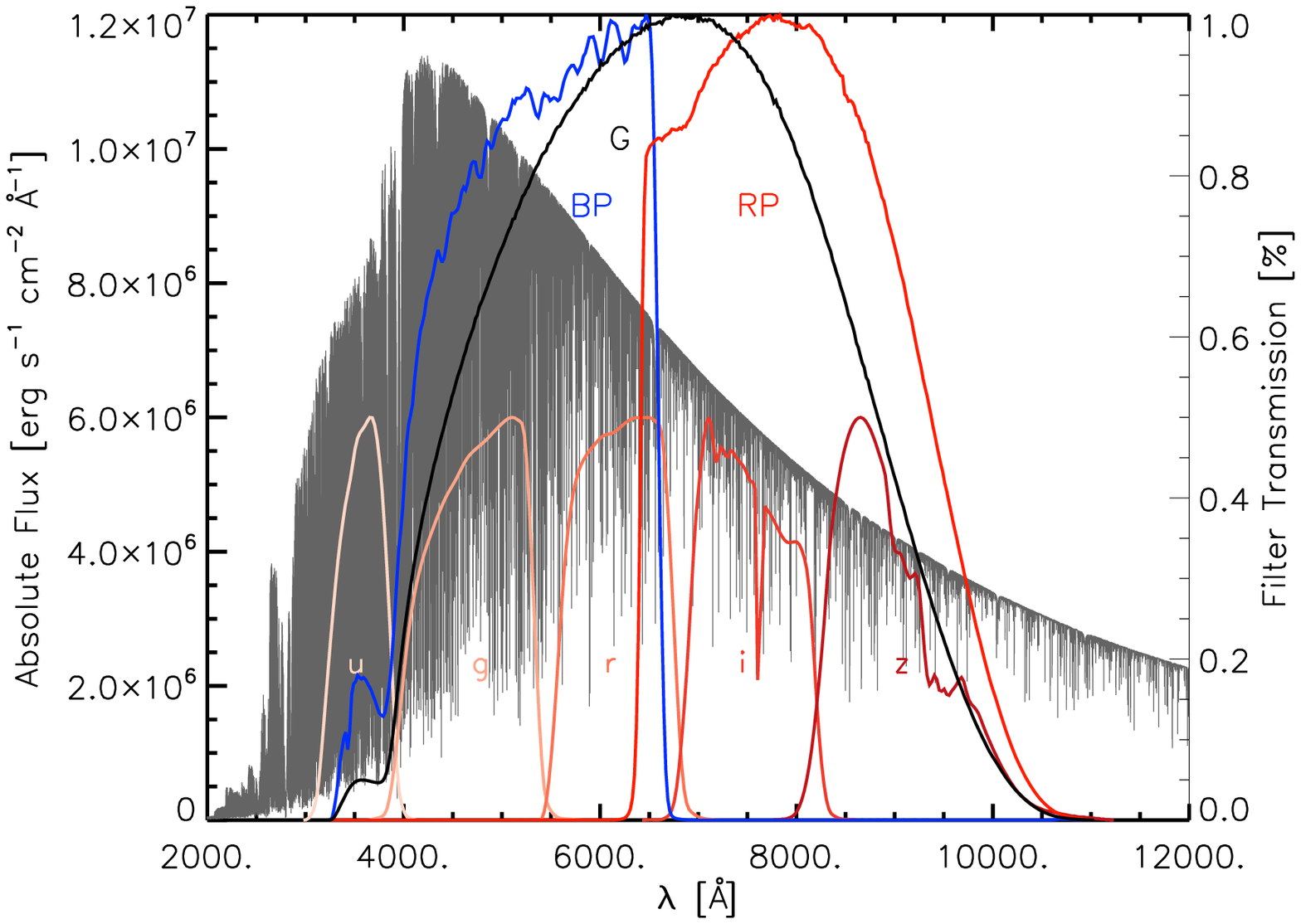} \\
      \includegraphics[width=0.47\hsize]{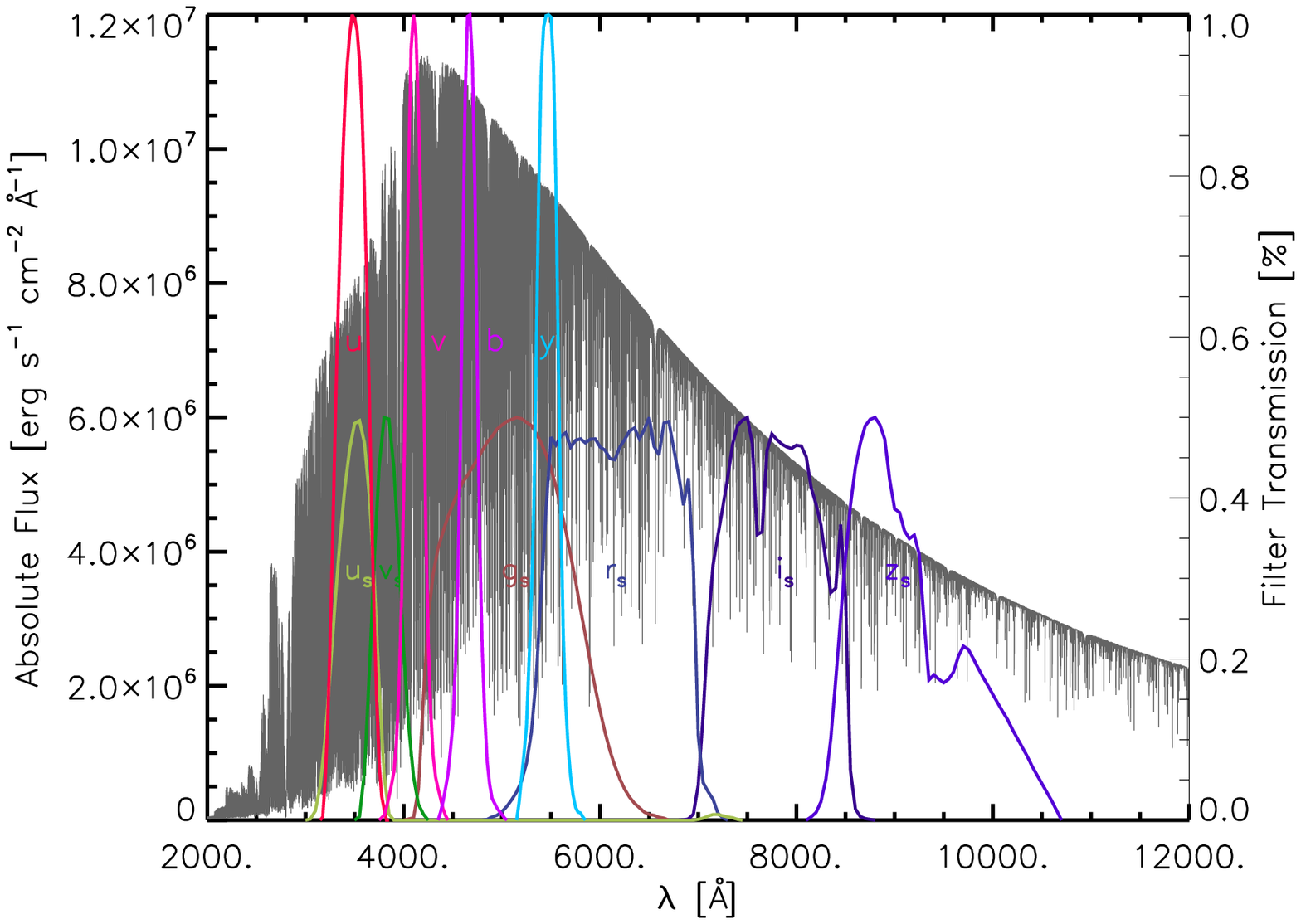} 
       \includegraphics[width=0.47\hsize]{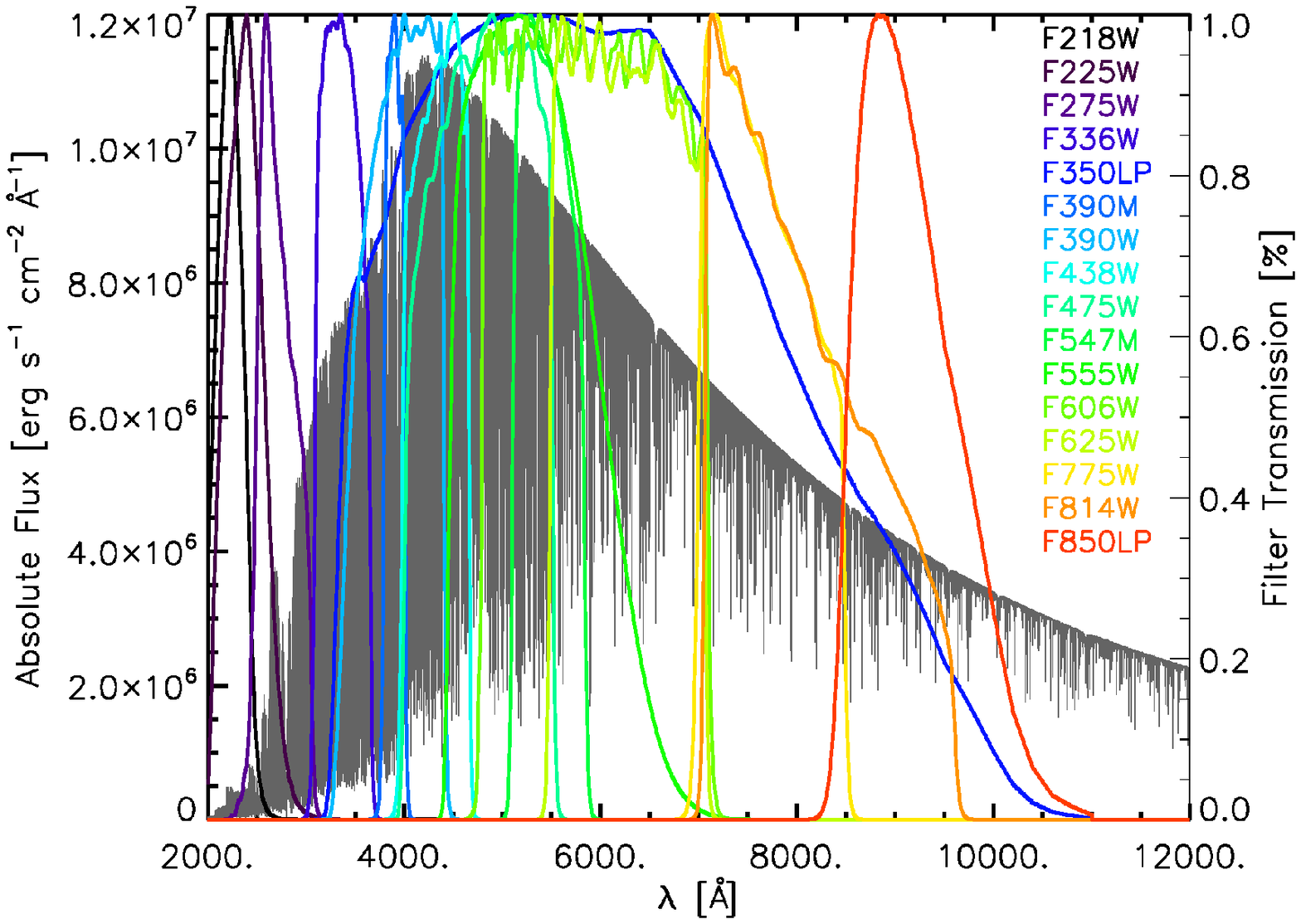} \\
  \end{tabular}
      \caption{Synthetic spectrum of the solar simulation at full spectral resolution in the spectral range 2000 -- 25\ 000 \AA\ (grey, top panel) and 2000 -- 12\ 000 \AA\ (central and bottom panels). Several system response functions (Table~\ref{tablefilters}), from which synthetic colours have been computed, are overplotted. Johnson-Cousins system response functions (U, B, V, Rc, Ic) are plotted in green  and 2MASS in  pink-violet  (top left panel); SDSS (u, g, r, i, z) in  yellow-red  and Gaia (BP, RP, G)  (top right panel); Str\"omgren (uvby) in  red-blue  and SkyMapper (u$_{\mathtt{s}}$, v$_{\mathtt{s}}$, g$_{\mathtt{s}}$, r$_{\mathtt{s}}$, i$_{\mathtt{s}}$, z$_{\mathtt{s}}$) (bottom left panel); and the 15 filters of the HST-WFC3 (bottom right panel). For clarity, SDSS and SkyMapper functions are normalised to 0.5.}
        \label{spectraexample}
           \end{figure*}   

 \begin{table}[!h]
 \scriptsize
\caption{Photometric systems used in this work and overplotted on the synthetic spectra in Fig.~\ref{spectraexample}.}             
\label{tablefilters}      
\centering                          
\renewcommand{\footnoterule}{} 
\begin{tabular}{c c  }        
\hline\hline                 
Photometric System name & Filters \\
            \hline
Johnson-Cousins\tablefootmark{a} & UBV(RI)$_{\rm{C}}$    \\
Sloan Digital Sky Survey (SDSS)\tablefootmark{b} & ugriz \\
2MASS\tablefootmark{c} & J, H, K$_{\rm{S}}$ \\
Gaia\tablefootmark{d} & G, BP, RP \\
SkyMapper\tablefootmark{e}  & u$_{\mathtt{s}}$, v$_{\mathtt{s}}$, g$_{\mathtt{s}}$, r$_{\mathtt{s}}$, i$_{\mathtt{s}}$, z$_{\mathtt{s}}$ \\
Str\"omgren\tablefootmark{f}  & uvby \\ 
HST-WFC3\tablefootmark{g} & F218W, F225W, F275W, \\
& F336W, F350W, F390M, F390W, \\
&  F438W, F475W, F547M, F555W, \\
&  F606W, F625W, F775W, F814W, F850LP\\
\hline\hline                          
\end{tabular}
\tablefoot{
\tablefoottext{a}{\cite{2012PASP..124..140B}}
\tablefoottext{b}{\cite{2010AJ....139.1628D}}
\tablefoottext{c}{\cite{2003AJ....126.1090C}}
\tablefoottext{d}{\cite{2010A&A...523A..48J}}
\tablefoottext{e}{\cite{2011PASP..123..789B}}
\tablefoottext{f}{\cite{2011PASP..123.1442B}}
\tablefoottext{g}{\cite{hst}}
}
\end{table}     
       

\subsection{Microturbulence}

The stellar surface convection produces a velocity field where the emerging spectral lines form. The Doppler broadening of these lines is a direct consequence of the velocity field in these crucial layers \citep{2000A&A...359..729A,2009LRSP....6....2N}. In traditional 1D models, this effect can be accounted for by the use of arbitrary micro- and macroturbulence parameters. Full 3D line formation calculations using 3D RHD simulations have demonstrated that in late-type stars the required non-thermal Doppler line broadening is fully included in the convection-related motions of the stellar atmosphere \citep[e.g. ][]{2007A&A...469..687C}. One-dimensional microturbulence represents the small-scale end of turbulent motions and is applied to the spectral line absorption coefficient. It affects  the strong lines to a greater extent, reducing their saturation, and to a lesser extent the widths of weak lines. For 1D-based SEDs,  microturbulence partly redistributes the flux in spectral regions probed by the photometric systems, in particular in regions crowded with lines towards the blue and the ultraviolet, and in filters with smaller wavelength coverages \citep{2014MNRAS.444..392C}. 

The values of the microturbulence parameters are usually determined by comparing synthetic and observed spectral line profiles and line strengths and often using a depth-independent value. For reference, a typical value for dwarfs and subgiants is around 1$-$1.5 km/s, which  increases to 2$-$2.5 km/s for stars on the red giant branch \citep[e.g. ][]{2001AJ....121.2159G}. A constant value of 2 km/s is usually assumed in large grids of synthetic stellar spectra \citep{2004astro.ph..5087C,2005ESASP.576..565B}. To compute our 1D hydrostatic comparison models, we explored different values of microturbulence: 0, 1, and 2 km/s. We found that there is no clear and no unique relation between  microturbulence and the stellar parameters, as reported by \cite{2014MNRAS.444..392C}. For clarity we adopted, as a guiding example, a value of 1 km/s when performing 1D calculations.

\subsection{Three-dimensional versus one-dimensional bolometric differences in correction}        
        
The figures in the Appendix display the bolometric corrections between 3D simulations and the corresponding 1D hydrostatic models with microturbulence = 1km/s. The values of $BC$ are reported in Table~4 to Table~8 for all the filters and, to retrieve the absolute colours,  Eqs.~(\ref{eq:bc1}) and~(\ref{eq:bc2}) should be used. Considering the SDSS, SkyMapper, 2MASS, and HST-WFC3 systems, the overall deviations are limited to small fraction (less than 5$\%$) from $BC_{\rm{r}}$ to $BC_{\rm{Ks}}$, but they increase to 10$\%$ for $BC_u$ and in $BC_g$ (SDSS) and for $BC_g sky$, $BC_v sky$, and $BC_u sky$ (SkyMapper) where the  optical  and line crowded region of the spectrum is probed with rather narrow filters (Fig.~\ref{spectraexample}, bottom); these differences decline with increasing effective temperature. This behaviour is more or less visible for all the photometric systems; there is no  clear correspondence of $\Delta BC$ with the other stellar parameters. 
On a broad scale, the wide infrared photometric systems like 2MASS ($BC_{\rm{H}}$ and $BC_{\rm{Ks}}$) and optical Gaia ($BC_{\rm{G}}$) display a noticeable offset with respect to $\Delta BC=0.0$. This is due to a redistribution of the spectral energy flux among the different filters and is a direct effect of the impact of 3D dynamics and thermodynamic structure on spectral line formation. \\
Gaia photometric systems return 3D and 1D deviations of less than 3$\%$ with higher values for the bluer system (BP). This effect may be not negligible and should become important for  future releases of Gaia data. For this purpose, part of the spectra presented in this work have already been provided to  Gaia consortium$-$CU8\footnote{For more details, see the technical note ``The 3D spectral library for BP/RP'' \citep{Gaia2}.}.

\section{Convective velocity shifts for RVS}
   
   Measurements of stellar radial velocities are fundamental in order to determine stellar space velocities. This is needed, for example to investigate the kinematic structure of stellar populations in the Galaxy or  to monitor for radial velocity variations, either of which would point to the presence of unseen companion(s). Convection plays a crucial role in the formation of spectral lines and deeply influences the shape, shift, and asymmetries of lines in late-type stars \citep[e.g. ][]{2000A&A...359..669A}. These stars represent most of the objects that will be observed during the Gaia mission. Absorption lines may be blueshifted as a result of convective movements in the stellar atmosphere: bright and rising convective elements contribute  more photons than the cool dark shrinking gas, and as a consequence, the absorption lines appear blueshifted \citep{1982ARA&A..20...61D}. However, the convective line shift is not the same for all the spectral lines. Each line has a unique fingerprint in the spectrum that depends on line strength, depth, shift, width, and asymmetry across the granulation pattern depending on their height of formation and sensitivity to the atmospheric conditions. In this context, the line strengths   play a major role \citep{2000A&A...359..743A}.
   
       \begin{figure}
   \centering
    \begin{tabular}{c}
       \includegraphics[width=0.99\hsize]{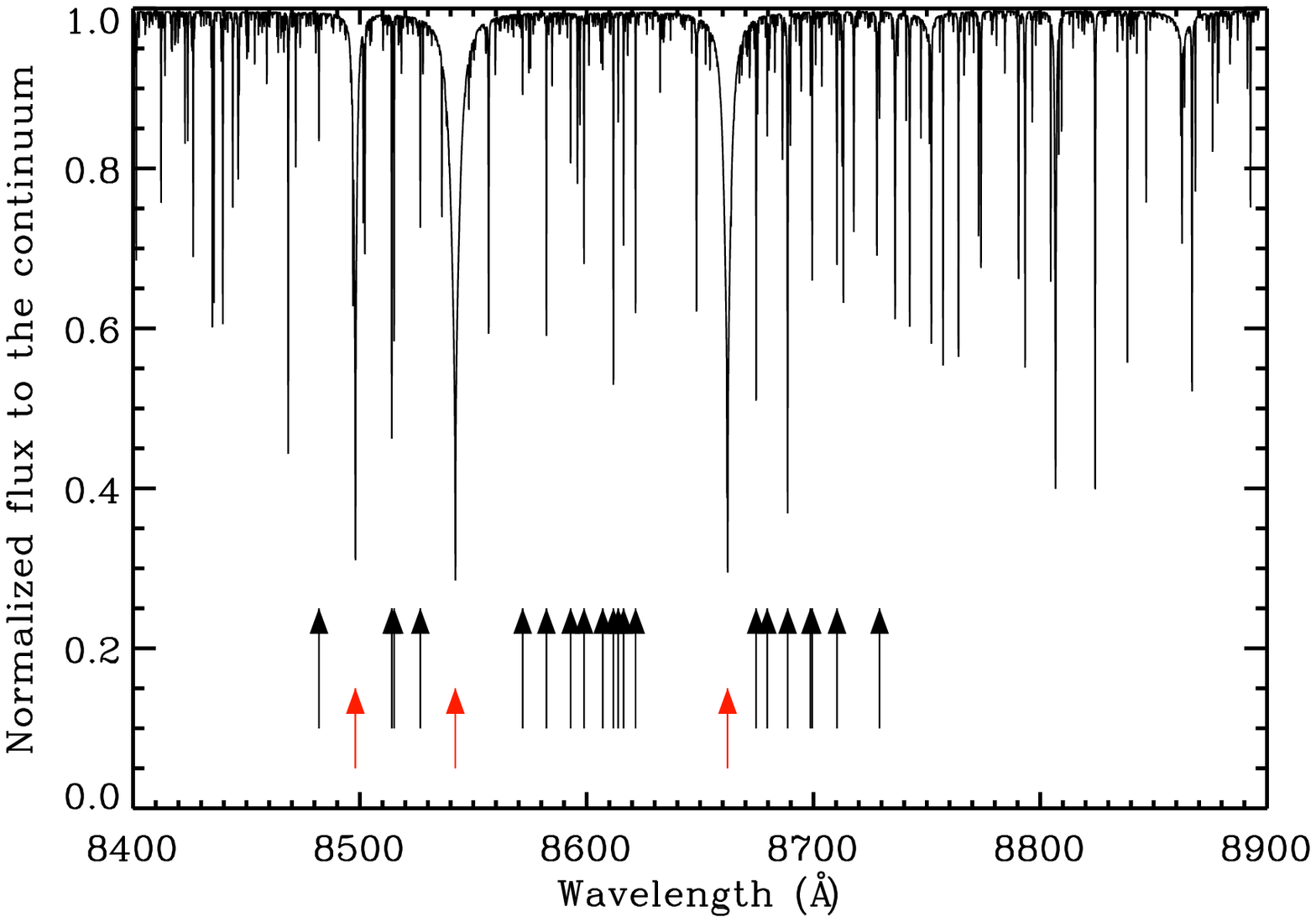} \\
  \end{tabular}
      \caption{Flux normalised to the continuum for the solar simulation (Table~4) in the RVS range (8400--8900 \AA\ ). The Fe $\mathrm{I}$ lines (black arrows) are from \cite{2006MNRAS.372..609B}, while the Ca II triplet are indicated with red arrows.}
        \label{gaia_rendering}
           \end{figure}  
           
   The aim of the present section is to derive the overall convective shift for 3D simulations. First, we computed the 1D and 3D spectra with a constant resolving power of $\lambda/\Delta\lambda=$300\ 000 from 8470 to 8710 \AA\  for a limited number of 3D simulations (see Table~4) covering stellar parameters observed by RVS (i.e. [Fe/H] $\geq$ -2.0). Then, from our spectra we selected only a series of non-blended Fe $\mathrm{I}$ lines and masked the others (Fig.~\ref{gaia_rendering}). The oscillator strengths of these Fe $\mathrm{I}$ lines (Table \ref{felines}) have been accurately determined by \cite{2006MNRAS.372..609B} using 3D RHD simulation where the Fe $\mathrm{I}$ and Ca II lines are indicated. It should be  noted that the synthetic spectra, when compared to the observations, have to be gravitationally redshifted \citep[e.g. ][]{2011A&A...526A.127P} by a certain amount corresponding to the type of star considered \citep[e.g. for the Sun it is  $636.486\pm0.024$ m/s][]{2003A&A...401.1185L}. Gravitational shifts for late-type dwarfs (log$g$$\approx4.5$) range between 0.7 and 0.8 km/s and they dramatically decrease with surface gravity down to 0.02-0.03 km/s for K giant stars with log$g$$\approx1.5$ \citep{2013A&A...550A.103A}.

  \begin{figure*}
  \begin{tabular}{cc}
   \includegraphics[width=0.5\hsize]{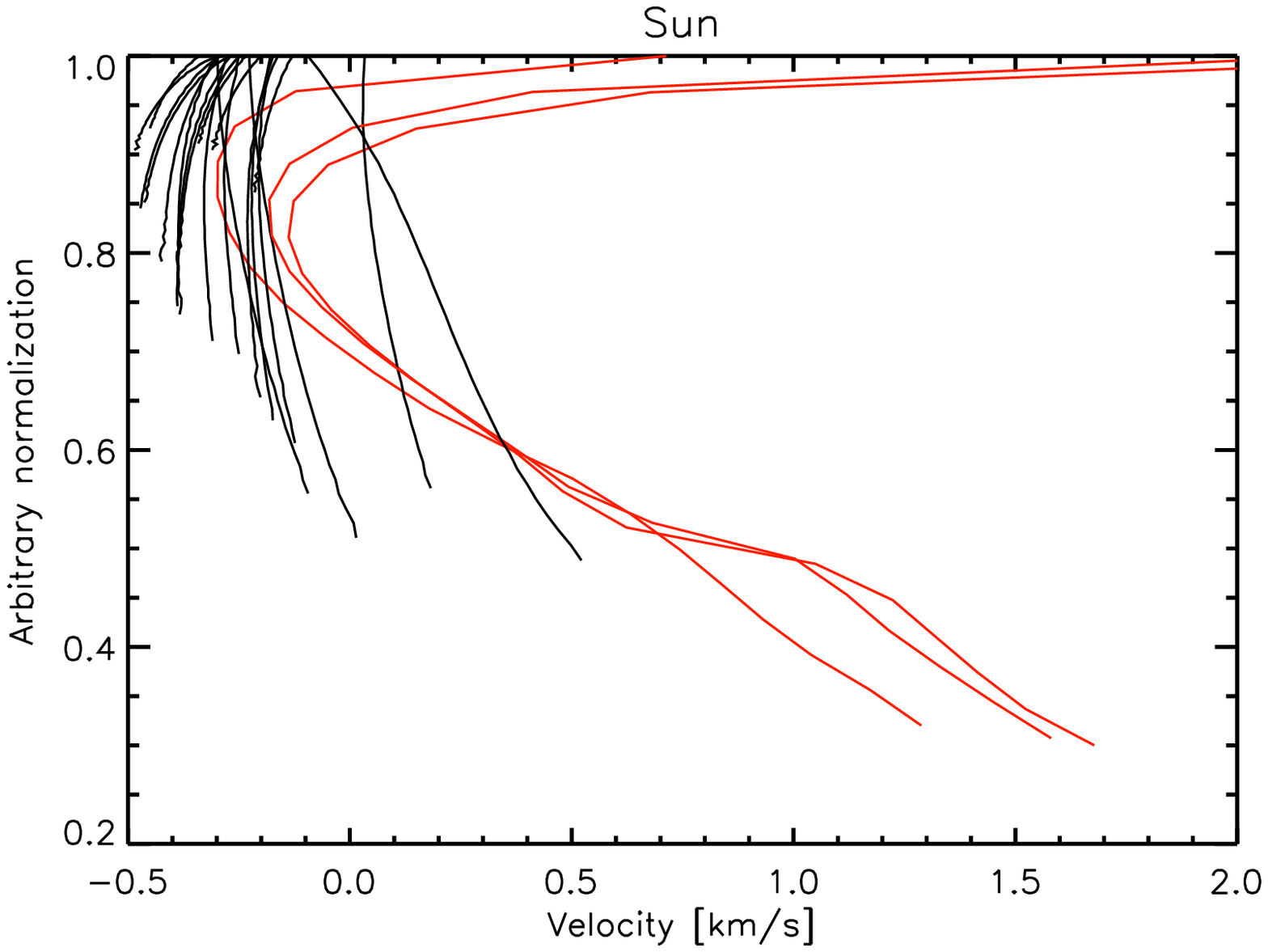}
   \includegraphics[width=0.5\hsize]{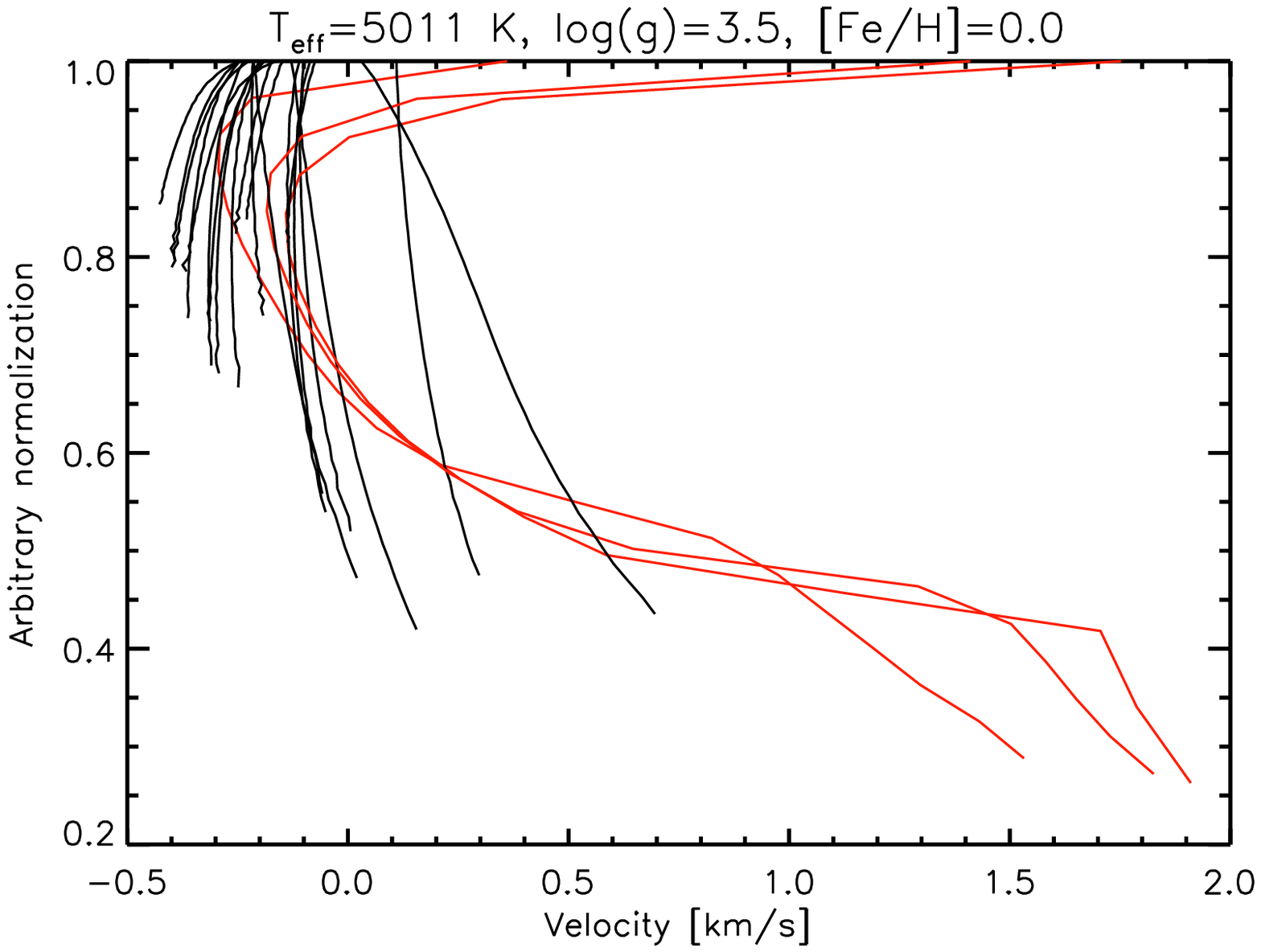}\\
   \includegraphics[width=0.5\hsize]{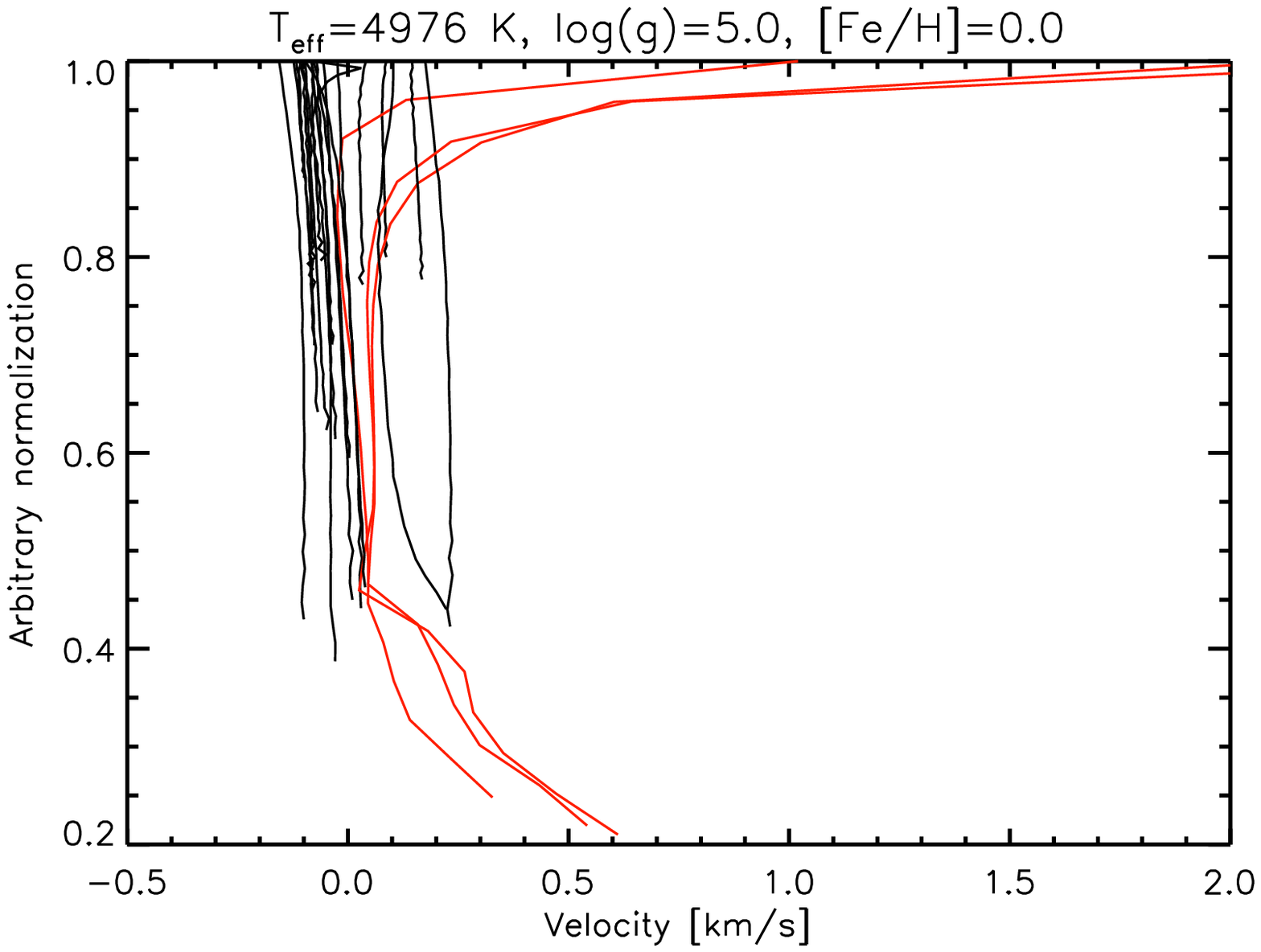}
   \includegraphics[width=0.5\hsize]{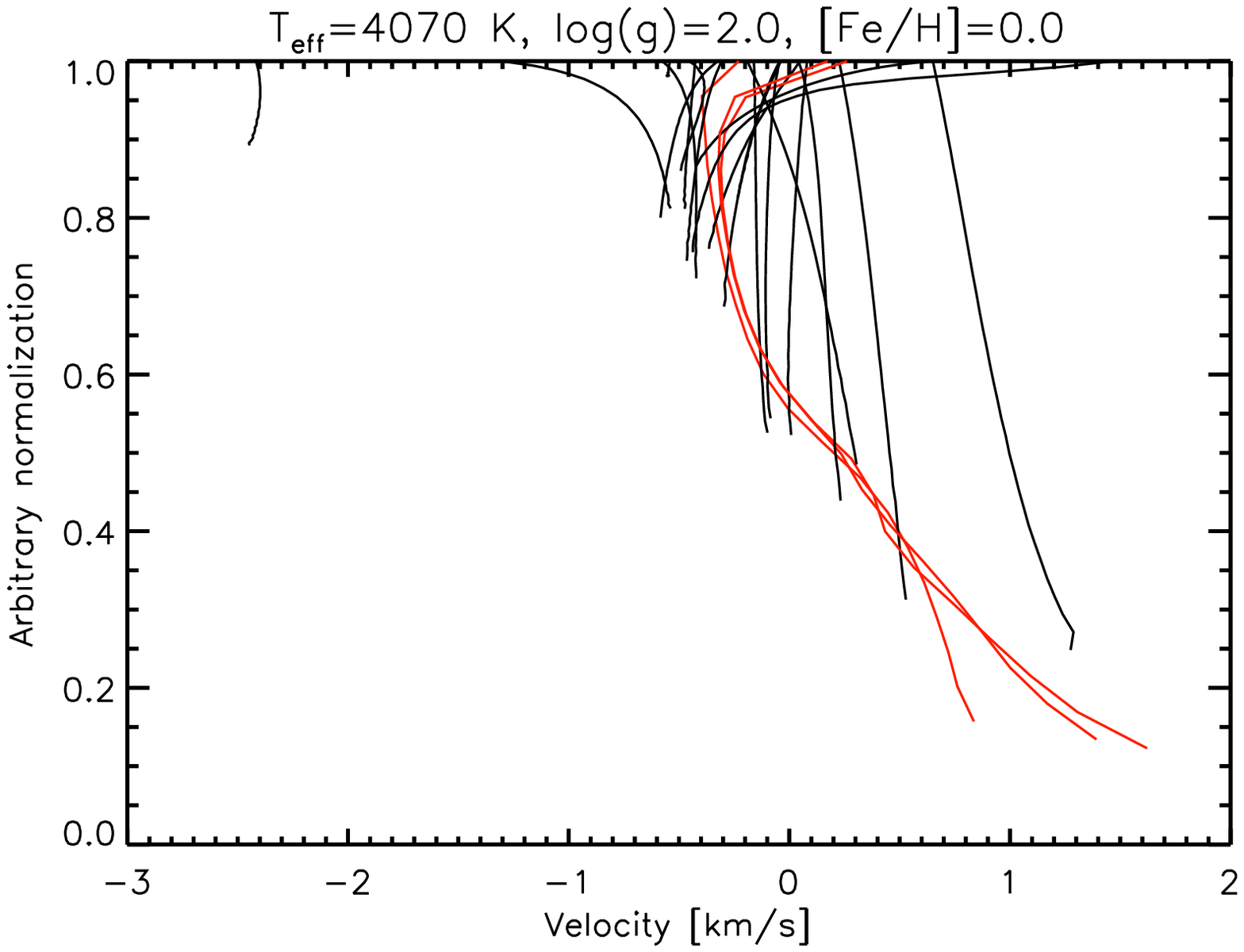}\\
   \includegraphics[width=0.5\hsize]{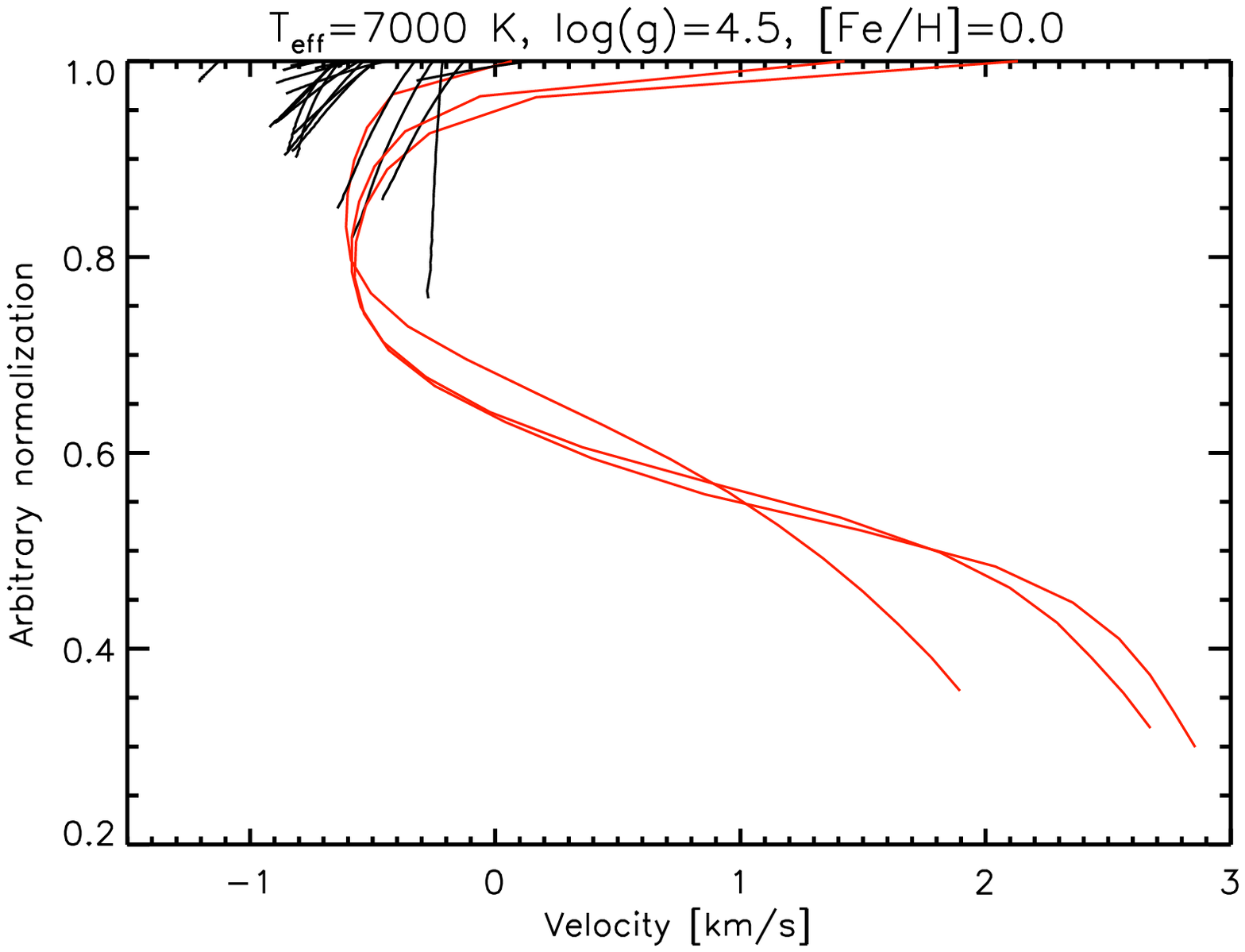}
\end{tabular}
      \caption{Line bisectors of the 20 Fe $\mathrm{I}$ lines (black) from \citep{2006MNRAS.372..609B} and Ca $\mathrm{II}$ triplet (red) for five 3D simulations in the grid.}
         \label{bisector}
   \end{figure*}
    
     \begin{figure*}
  \begin{tabular}{cc}
   \includegraphics[width=0.48\hsize]{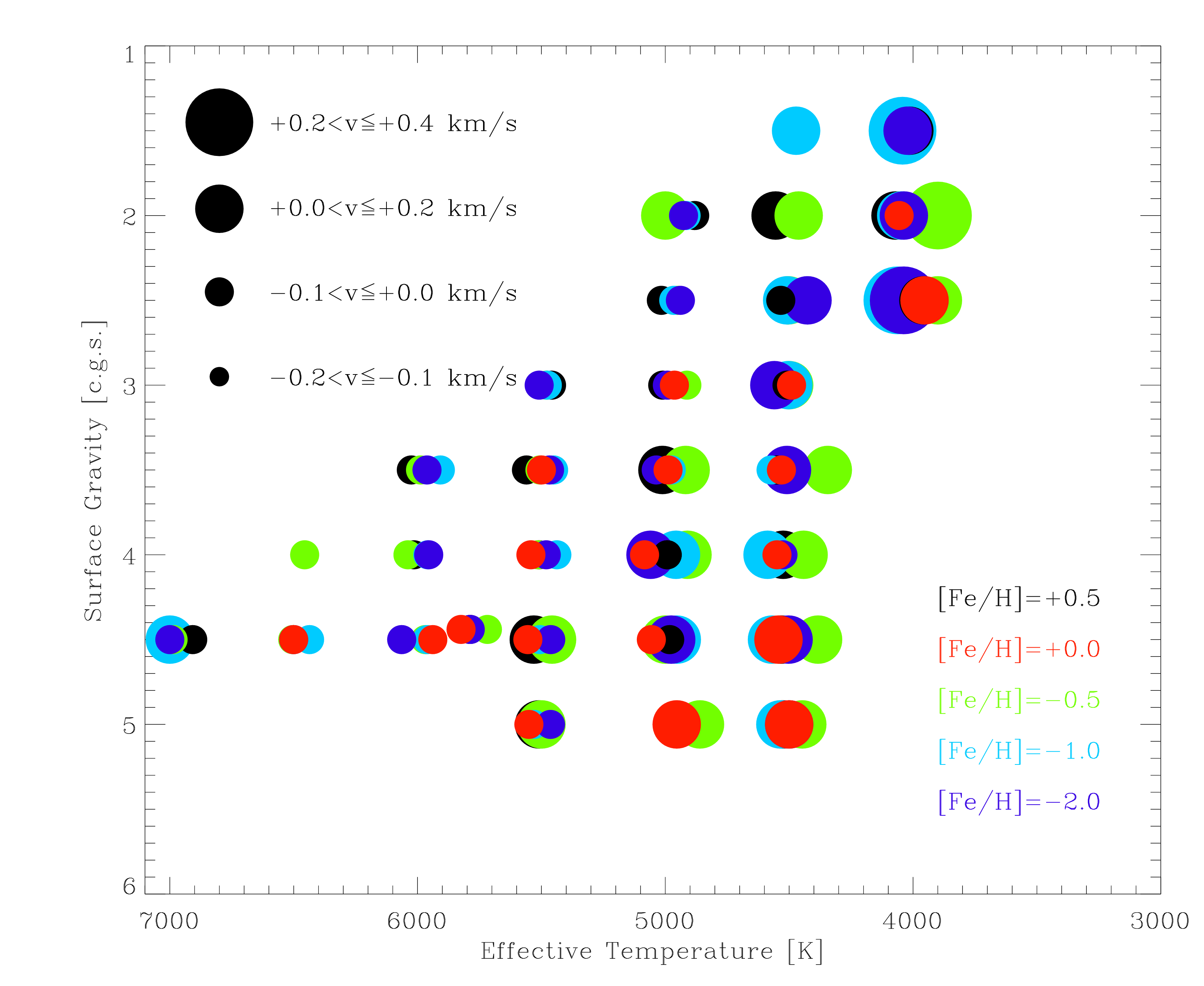}
   \includegraphics[width=0.48\hsize]{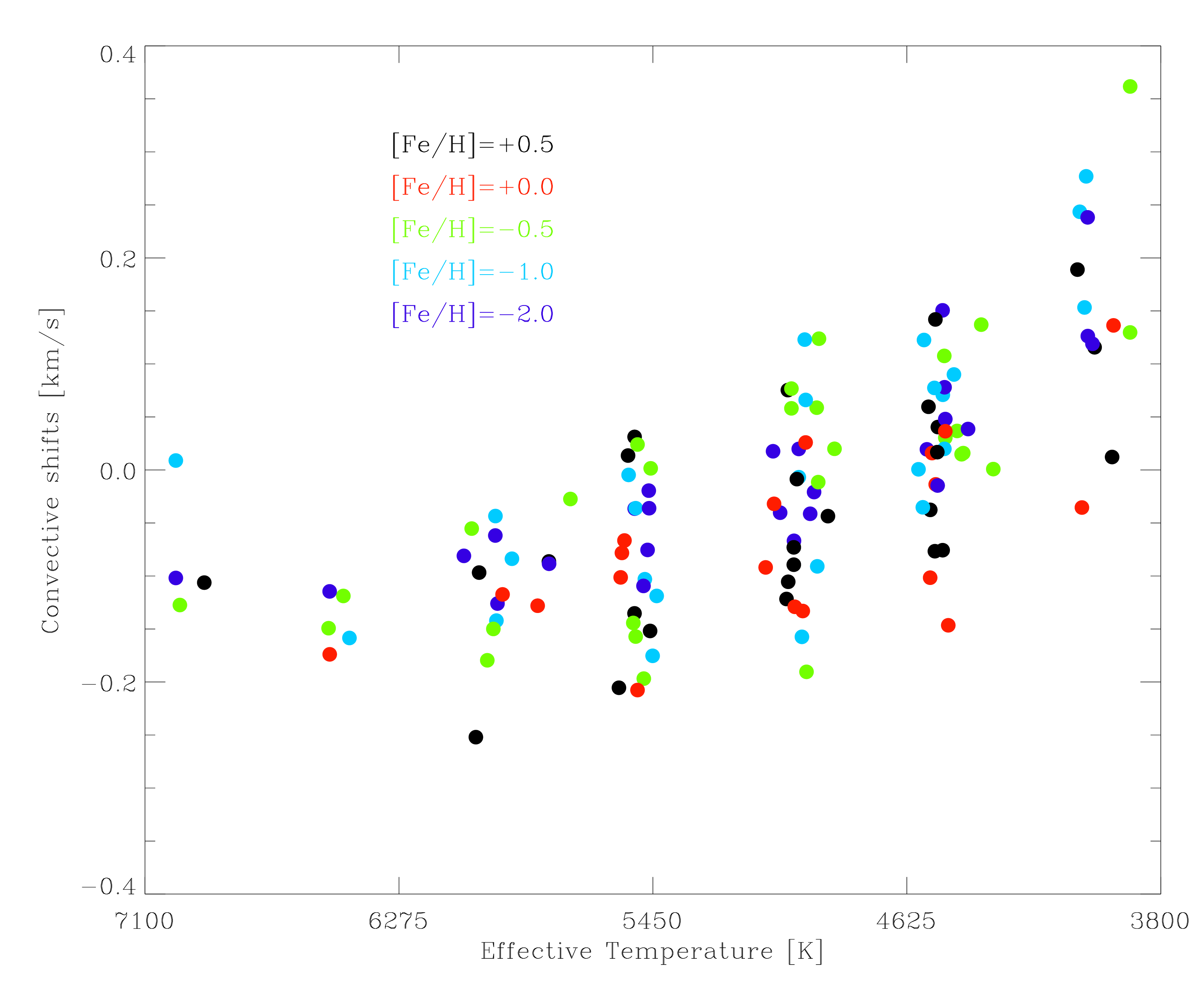} \\
    \includegraphics[width=0.48\hsize]{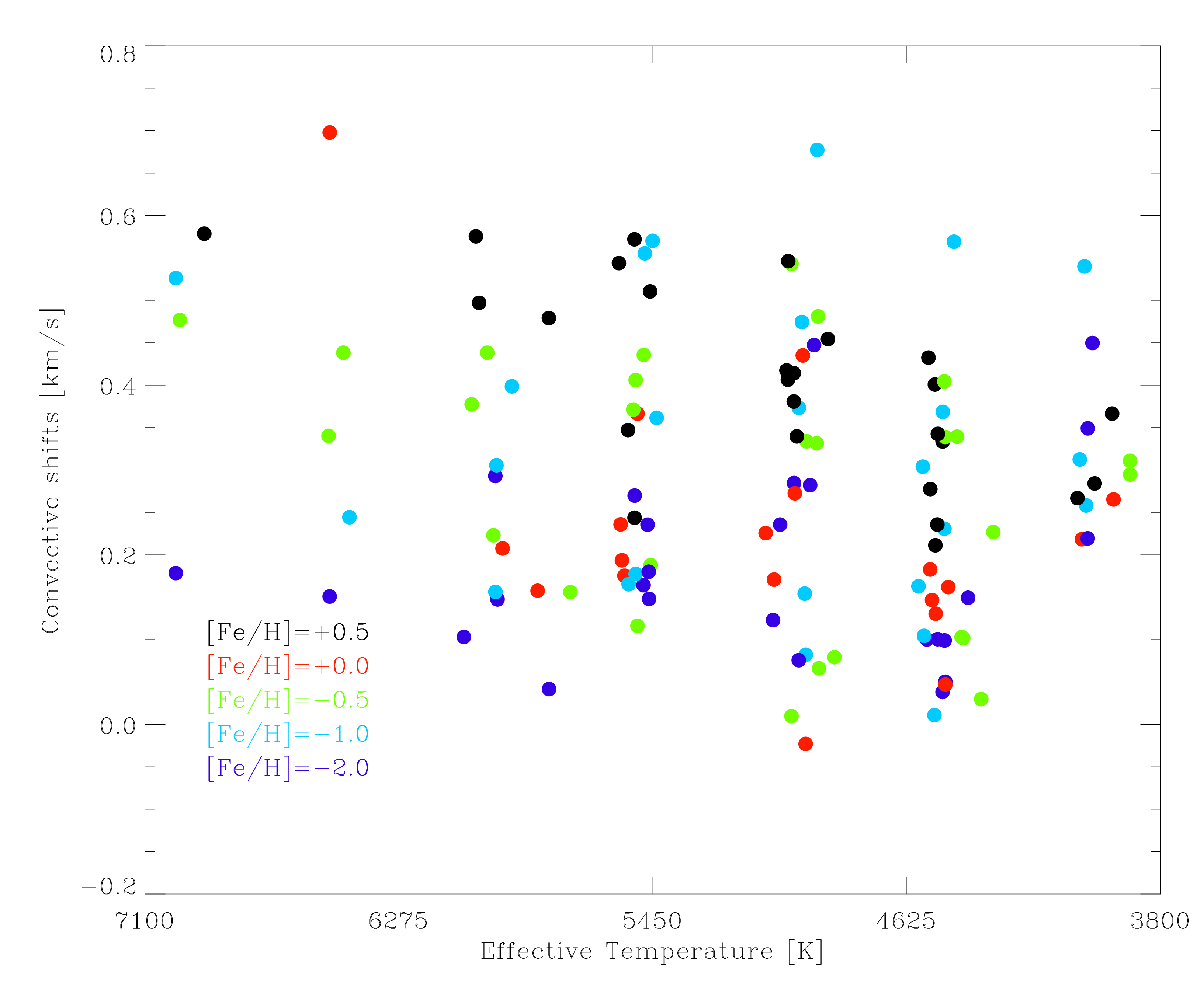} 
   \includegraphics[width=0.48\hsize]{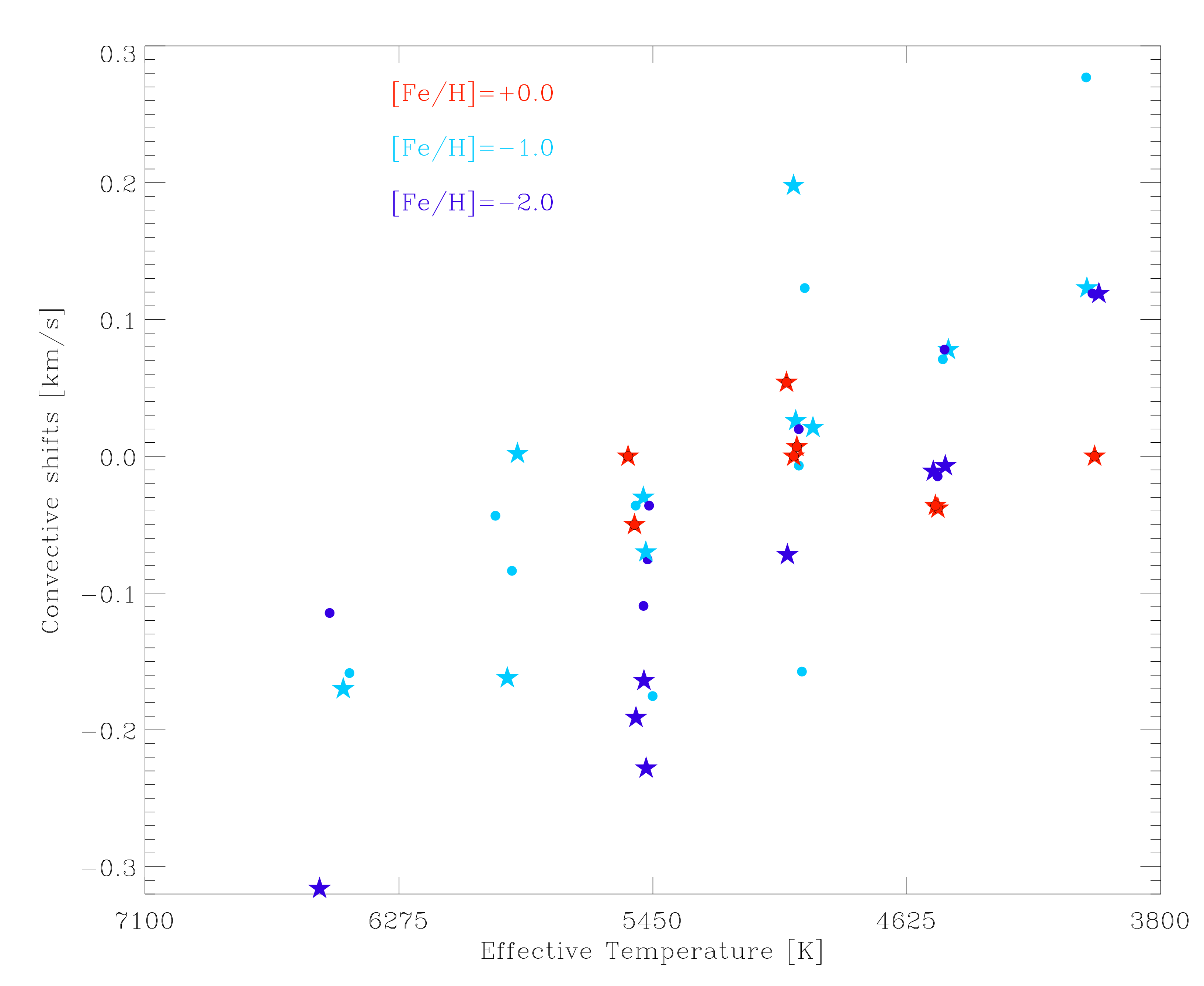}
\end{tabular}
      \caption{\emph{Top left panel:} Convective shifts predicted by the 3D hydrodynamical simulations for the Gaia RVS spectral range and the Fe $\mathrm{I}$ lines (see text for details). \emph{Top right panel:} Convective shifts from Fe $\mathrm{I}$ lines as a function of the effective temperature of the 3D simulations in Table~4. \emph{Bottom left panel:} Convective shifts from Ca II lines. \emph{Bottom right panel:} Comparison of convective shifts  for a selected number of RHD simulations from this work (stars) with simulations with equivalent stellar parameters from \cite{2013A&A...550A.103A} (circles). See Table~\ref{differencesallende} for details.}
         \label{centroid}
   \end{figure*}

    \begin{table}
     
\caption{Central wavelength position, oscillator strength (log $gf$), and excitation potential ($\chi$) for the 20 Fe $\mathrm{I}$ in the spectral domain of RVS \citep{2006MNRAS.372..609B}}
\label{felines}      
\centering                          
\renewcommand{\footnoterule}{} 
\begin{tabular}{c c c}        
\hline\hline                 
$\lambda$ [\AA] & (log $gf$) & $\chi$ [eV] \\
\hline
8481.985   &   -2.097 & 4.1860  \\
8514.068   &   -2.250 & 2.1980 \\
8515.109   &    -2.033 & 3.0180 \\
8526.667   &    -0.675 & 4.9130 \\
8571.803 & -1.134  &   5.0100 \\
8582.257 & -2.198  &   2.9900 \\
8592.951 & -0.891  &  4.9560  \\
8598.829 & -1.285  & 4.3860 \\
8607.078 & -1.419  &   5.0100 \\
8611.801 & -1.900  &   2.8450 \\
8613.939 & -1.121 & 4.9880 \\
8616.280 & -0.935  &   4.9130 \\
8621.601 & -2.369  &   2.9490 \\
8674.741 & -1.780  &  2.8310  \\
8679.639 & -1.040  & 4.9660 \\
8688.623 & -1.249  &  2.1760   \\
8698.706 & -3.464  &  2.9900    \\
8699.453 & -0.480  &  4.9550   \\
8710.391 & -0.425  &  4.9130   \\
8729.147 & -2.933  &  3.4150  \\
\hline\hline                          
\end{tabular}
\end{table}

The velocity gradient through the photosphere sets the basic shape of the absorption lines in terms of asymmetry and position of the emerging intensity.  One way to detect the asymmetries in the line is the bisector\footnote{Defined as the locus of the midpoints of the spectral line}. A symmetric profile has a straight vertical bisector (i.e. in the case of hydrostatic 1D spectra). The spectral lines with C-shape bisectors are formed mostly in the upflows (granules) and therefore blueshifted. The reverse C-shapes are generally formed in downflows \citep{1981A&A....96..345D}. Reversed C-shape bisectors can be explained by a combination of a steep decline in velocities with height with a flux deficit spanning only a fraction of the red wing of the line profiles \citep{2010ApJ...721..670G}. Different articles show the presence of bisectors revealing asymmetries and wavelength shifts that indicate the presence of granulation for several kinds of stars \citep[e.g. ][]{2008A&A...492..841R,2009ApJ...697.1032G}. \\
Figure~\ref{bisector} shows the line bisectors for the Fe $\mathrm{I}$ and Ca II triplet lines for stars with different $T_{\rm{eff}}$ and log$g$, but with the same metallicity. The gas is strongly horizontally divergent due to mass conservation and its velocities diminish with height. Weak lines (with typically high excitation potential), which form in deeper layers, are more blueshifted than strong lines whose core and part of the wings are formed in higher layers. This effect is particularly visible in Fig.~\ref{bisector} when comparing the solar bisector with the hottest $T_{\rm{eff}}=7000$K simulations. In addition, the velocity field in 3D simulations of \textsc{Stagger}-grid largely affects the overall shape of the iron lines in the range of RVS and for all the stars with the strongest effects for Ca II. This has already been shown for other spectral regions \citep[e.g. ][]{2000A&A...359..729A,2002ApJ...567..544A,2009A&A...501.1087R,2013A&A...554A.118P,2014arXiv1403.6245M}. \\
We determined the convective shift considering only Fe $\mathrm{I}$ and only Ca II triplet lines, we cross-correlated each 3D spectrum with the corresponding 1D by using a lag vector corresponding to radial velocities (RV) in the range $-$10 < $v$ < $+$10 km/s for Fe $\mathrm{I}$, and $-$150 < $v$ < $+$150 km/s for Ca II triplet lines, in steps of 0.3 km/s. These velocity ranges were chosen to largely cover the wavelength frequency points of all the single lines. For each RV value, we Doppler-shifted the 1D spectrum and computed its cross-correlation function (CCF) with the 3D spectrum. The final step is to compute the weighted average to obtain the location of the CCF maximum, which corresponds to the actual 3D convective shifts ($CS$) with respect to 1D models: 

\begin{equation}
CS = \frac{\int^{+10}_{-10}RV\left(v\right)\cdot CCF\left(v\right)dv}{\int^{+10}_{-10}CCF\left(v\right)dv}
 \end{equation}

  \begin{table}
    \scriptsize 
\caption{Selected convective shifts for Fe $\mathrm{I}$ only ($CS_{Stagger}$ in km/s) from RHD simulations in this work and from CIFIST-grid simulations ($CS_{CIFIST}$ in km/s) with equivalent stellar parameters from \cite{2013A&A...550A.103A}. The difference in $T_{\rm{eff}}$ is set to be smaller than 50 K. }
\label{differencesallende}      
\centering                          
\renewcommand{\footnoterule}{} 
\begin{tabular}{c c c c c c c}        
\hline\hline                 
$T_{\rm{eff, Stagger}}$ & $T_{\rm{eff, CIFIST}}$ & log$g$ & [Fe/H] & $CS_{Stagger}$ & $CS_{CIFIST}$ & $\Delta_{CS}$  \\
  \hline
        4014   &     4018 &  1.50 &  -0.0 &   0.116 &   0.000  &  0.116     \\
        4524   &     4480 &  4.00 &  -0.0 &   0.040 &  -0.038  &  0.078    \\
        4532   &     4509 &  4.50 &  -0.0 &   0.142 &  -0.036  &  0.178    \\
        5015   &     4968 &  2.50 &  -0.0 &  -0.122 &   0.054  & -0.176   \\
        4992   &     4954 &  4.00 &  -0.0 &  -0.073 &   0.000  & -0.073    \\
        4982   &     4982 &  4.50 &  -0.0 &  -0.009 &   0.007  & -0.016    \\
        5509   &     5475 &  4.00 &  -0.0 &  -0.135 &  -0.050  & -0.085   \\
        5530   &     5488 &  4.50 &  -0.0 &   0.014 &   0.000  &  0.014   \\
        4042   &     4040 &  1.50 &  -1.0 &   0.277 &   0.123  &  0.154   \\
        4508   &     4490 &  2.50 &  -1.0 &   0.071 &   0.078  & -0.007  \\
        4965   &     4993 &  2.50 &  -1.0 &  -0.157 &   0.198  & -0.355  \\
        4975   &     4930 &  3.50 &  -1.0 &  -0.007 &   0.021  & -0.028   \\
        4956   &     4986 &  4.00 &  -1.0 &   0.123 &   0.026  &  0.097   \\
        5450   &     5481 &  3.50 &  -1.0 &  -0.175 &  -0.030  & -0.145  \\
        5506   &     5473 &  4.50 &  -1.0 &  -0.036 &  -0.070  &  0.034   \\
        5907   &     5890 &  3.50 &  -1.0 &  -0.084 &   0.002  & -0.086   \\
        5961   &     5923 &  4.50 &  -1.0 &  -0.043 &  -0.162  &  0.119   \\
        6435   &     6456 &  4.50 &  -1.0 &  -0.158 &  -0.170  &  0.012   \\
        4021   &     4001 &  1.50 &  -2.0 &   0.119 &   0.119  & -0.000    \\
        4524   &     4500 &  4.00 &  -2.0 &  -0.015 &  -0.007  & -0.008   \\
        4502   &     4539 &  4.50 &  -2.0 &   0.078 &  -0.011  &  0.089   \\
        4976   &     5013 &  4.50 &  -2.0 &   0.020 &  -0.072  &  0.092   \\
        5467   &     5505 &  3.50 &  -2.0 &  -0.075 &  -0.191  &  0.116   \\
        5480   &     5472 &  4.00 &  -2.0 &  -0.109 &  -0.228  &  0.119   \\
        5462   &     5479 &  4.50 &  -2.0 &  -0.036 &  -0.164  &  0.128   \\
        6500   &     6533 &  4.50 &  -2.0 &  -0.115 &  -0.316  &  0.201   \\ 
      \hline\hline                          
\end{tabular}
\end{table}     

Figure~\ref{centroid} displays the convective shifts for all the simulations either in the HR diagram (top left panel) or as a function of the $T_{\rm{eff}}$ for the Fe $\mathrm{I}$ of Table~\ref{felines} (top right panel), and  
for the Ca II triplet lines (bottom left panel). We found that surface gravity and metallicity have a small effect on the convective shifts, as already noticed by \cite{2013A&A...550A.103A}. The  values for the Fe $\mathrm{I}$ are in the range between  -0.235 and +0.361 km/s. The convective shifts of Ca II lines are strongly redshifted (as shown by red bisectors in Fig.~\ref{bisector}) and are between -0.023 and +0.698 km/s. In  Fig.~\ref{centroid}  (top right panel), there is a net correlation of the convective shifts with the effective temperature: $T_{\rm{eff}}\lessapprox4500$K denotes redshifts, while $T_{\rm{eff}}\gtrapprox5000$K denotes blueshifts (except for the hottest $T_{\rm{eff}}\approx7000$K). This result is in agreement with \cite{2013A&A...550A.103A}, who had performed the calculations for a different set of iron lines and found a milder correlation, where  $T_{\rm{eff}}$ with warmer stars tend to exhibit larger blueshifts. \\
To quantify the differences in the convective shifts, we selected 26 simulations from Table~4 with the same surface gravity, metallicity, and $\Delta T_{\rm{eff}}<50$K with respect to a subset of CIFIST-grid simulations from \cite{2013A&A...550A.103A}. Convective shifts, as a function of metallicity (Table~\ref{differencesallende}), from RHD simulations in this work are $CS_{Stagger,[Fe/H]=0}$=[-0.135, 0.142], $CS_{Stagger,[Fe/H]=-1}$=[-0.175, 0.277], and $CS_{Stagger,[Fe/H]=-2}$=[-0.114, 0.119] km/s, and from CIFIST-grid simulations $CS_{CIFIST,[Fe/H]=0}$=[-0.050, 0.054], $CS_{CIFIST,[Fe/H]=-1}$=[-0.170, 0.198], and $CS_{CIFIST,[Fe/H]=-2}$=[-0.316, 0.119] km/s. The spanned shift values from both grids are similar (Fig.~\ref{centroid}, bottom right panel), and show  smaller deviations at solar metallicity ($\Delta_{CS}\leqslant0.195$ km/s) and slightly larger deviations at [Fe/H]=-1 ($\Delta_{CS}\leqslant0.370$ km/s) and [Fe/H]=-2 ($\Delta_{CS}\leqslant0.221$ km/s).  Apart from the possible numerical differences  in the simulations and in the radiative transfer, the shift deviations may  also be due to the set of spectral lines considered.

The extraction of accurate radial velocities from RVS needs an appropriate wavelength calibration from convective shifts. This is directly processed in RVS pipeline using the synthetic spectra presented in this work and provided to  Gaia consortium$-$CU6\footnote{For more details, see the technical note ``3D spectral library for RVS radial velocities'' \citep{Gaia1} and the paper on CU6 design and performance  (Sartoretti, Katz et al., in preparation).}.

   \section{Conclusions}
   
   We provided synthetic spectra from the \textsc{Stagger}-grid:
   \begin{itemize}
   \item low-resolution  spectra from 1000 to 200\ 000 \AA\ with a constant resolving power of $\lambda/\Delta\lambda=$20\ 000;
   \item high-resolution spectra from 8400 to 8900 \AA\ (Gaia RVS spectral range), with a constant resolving power of $\lambda/\Delta\lambda=$300\ 000.
   \end {itemize}
   
We used the low-resolution spectra to compute synthetic colours in the Johnson-Cousins $UBV(RI)_C$, SDSS, 2MASS, Gaia systems, SkyMapper, Str\"omgren, and HST-WFC3. We extracted the bolometric corrections for the  3D simulations and the corresponding 1D hydrostatic models. We probed that 1D versus 3D deviations are limited to small fraction (less than 5$\%$) except for $BC_u$ and to a lesser extent $BC_g$, where the differences are larger than 10$\%$. Systems u and g  span the optical and line crowded region of the spectrum. Moreover, we showed that there is a clear correlation between effective temperature and 3D and 1D deviations ($\Delta BC$): it decreases with increasing  effective temperature.\\
The Gaia photometric system return 3D and 1D deviations of less than 3$\%$ with higher values for the bluer filter (BP). This effect should become important for future releases  of Gaia data. For this purpose, part of the spectra presented in this work have already been provided to  Gaia consortium$-$CU8.

We used the high-resolution spectra to denote the impact of the velocity gradient through the photosphere on the basic shape of the absorption lines. For this purpose, we reported the line bisectors of non-blended Fe $\mathrm{I}$ and Ca II triplet lines for different stars. We showed that weak lines (high excitation potential), which form in deeper layers, are more blueshifted than strong lines (low excitation potential), whose core and part of the wings are formed in higher layers. \\
As a final step to derive the overall convective shift for 3D simulations with respect to the reference 1D hydrostatic models, we cross-correlated each 3D spectrum with the corresponding 1D spectrum. The spanned values are between -0.235 and +0.361 km/s. We showed a net correlation of the convective shifts with the effective temperature: lower $T_{\rm{eff}}$ denotes redshifts and higher $T_{\rm{eff}}$ blueshifts; this result is in agreement with \cite{2013A&A...550A.103A}. In addition, we quantified the differences in the convective shifts between a subset of the RHD simulations in this work and the corresponding CIFIST-grid simulations. The spanned shift values from the two grids are similar, and show  smaller deviations at solar metallicity. The extraction of accurate radial velocities from RVS spectra need an appropriate wavelength calibration from convection shifts. The spectra presented in this work have been provided to  Gaia consortium$-$CU6 to directly process the observed spectra in RVS pipeline.

  We have made all the spectra publicly available for the community through the POLLUX database \citep{2010A&A...516A..13P}. POLLUX\footnote{Available at http://pollux.oreme.org} is a stellar spectra database proposing access to theoretical data including high-resolution synthetic spectra and spectral energy distributions from several model atmospheres. Continuous development either of the \textsc{Stagger}-grid simulations or of the spectral synthesis calculations will be uploaded there in the future.

\longtab{
\scriptsize
\begin{landscape}
\label{simus1}
\begin{longtable}{c c c | c | c c c c c | c c c | c c c c c | c c c}
\caption{RHD simulations' stellar parameters (first three columns), bolometric Magnitude (M$_{\rm{bol}}$), and bolometric correction ($BC$) for Johnson-Cousins, 2MASS, SDSS, and Gaia systems (Table~\ref{tablefilters}).  }\\
\hline\hline
$T_{\rm{eff}}$ & log$g$ & [Fe/H] & $M_{\rm{bol}}$ & $BC_{\rm{U}}$ & $BC_{\rm{B}}$ & $BC_{\rm{V}}$ & $BC_{\rm{R}}$ & $BC_{\rm{I}}$ & $BC_{\rm{J}}$ & $BC_{\rm{H}}$ & $BC_{\rm{Ks}}$ & $BC_{\rm{u}}$ & $BC_{\rm{g}}$ & $BC_{\rm{r}}$ & $BC_{\rm{i}}$ & $BC_{\rm{z}}$ & $BC_{\rm{G}}$ & $BC_{\rm{BP}}$ & $BC_{\rm{RP}}$ \\
\hline
4014.99 & 1.50 &  0.0 & 6.320 &   -3.75310 &   -2.26910 &   -0.90560 &   -0.14050 &    0.53170 &    1.58810 &    2.32330 &    2.53140 &   -4.67960 &   -1.63880 &   -0.43510 &    0.02130 &    0.31350 &   -0.42180 &   -1.17700 &    0.44010 \\ 
4042.38 & 1.50 & -1.0 & 6.291 &   -3.11690 &   -2.12650 &   -0.86970 &   -0.13680 &    0.52720 &    1.55730 &    2.32870 &    2.50310 &   -4.01150 &   -1.54990 &   -0.42970 &    0.02220 &    0.30510 &   -0.40820 &   -1.13330 &    0.43660 \\ 
4021.90 & 1.50 & -2.0 & 6.313 &   -3.24590 &   -2.23800 &   -0.90050 &   -0.14700 &    0.53780 &    1.56940 &    2.32730 &    2.50270 &   -4.17080 &   -1.62720 &   -0.44840 &    0.02840 &    0.32220 &   -0.42050 &   -1.17860 &    0.44480 \\ 
3951.92 & 1.50 & -3.0 & 6.389 &   -3.90090 &   -2.60760 &   -1.04190 &   -0.19200 &    0.56190 &    1.61550 &    2.33340 &    2.52470 &   -4.86820 &   -1.90680 &   -0.52040 &    0.03670 &    0.36450 &   -0.46890 &   -1.33700 &    0.46060 \\ 
4070.69 & 2.00 &  0.0 & 6.260 &   -3.47470 &   -2.14830 &   -0.85400 &   -0.10740 &    0.54590 &    1.57170 &    2.29980 &    2.49240 &   -4.39490 &   -1.55290 &   -0.39570 &    0.04140 &    0.31920 &   -0.38850 &   -1.11670 &    0.45590 \\
\end{longtable}
\end{landscape}
}

\longtab{
\begin{landscape}
\label{simus2}
\begin{longtable}{c c c | c | c c c c c c | c c c }
\caption{Same as in Table~4 with bolometric correction for SkyMapper photometric system of Table~\ref{tablefilters}. The last three columns are the Str\"omgren index $b-y$, $m1 = (v-b)-(b-y)$, and 
$c1 = (u-v)-(v-b)$ \citep{2005ARA&A..43..293B}.}\\
\hline\hline
$T_{\rm{eff}}$ & log$g$ & [Fe/H] & $M_{\rm{bol}}$ & $BC_{\rm{us}}$ & $BC_{\rm{vs}}$& $BC_{\rm{gs}}$ & $BC_{\rm{rs}}$ & $BC_{\rm{is}}$ & $BC_{\rm{zs}}$ &  $b-y$ & $m1$ &  $c1$ \\
\hline
4014.99 & 1.50 &  0.0 & 6.320 &   -4.50340 &   -3.96180 &   -1.21350 &   -0.46120 &    0.08130 &    0.33140 &    0.86270 &    0.54200 &    0.54670 \\ 
4042.38 & 1.50 & -1.0 & 6.291 &   -3.95760 &   -3.40670 &   -1.16070 &   -0.45220 &    0.08050 &    0.31940 &    0.80280 &    0.40080 &    0.37820 \\ 
4021.90 & 1.50 & -2.0 & 6.313 &   -4.13890 &   -3.45350 &   -1.21800 &   -0.47130 &    0.08910 &    0.33620 &    0.88560 &    0.33970 &    0.44090 \\ 
3951.92 & 1.50 & -3.0 & 6.389 &   -4.72750 &   -4.00560 &   -1.42040 &   -0.54570 &    0.10710 &    0.38030 &    1.08410 &    0.28450 &    0.58400 \\ 
4070.69 & 2.00 &  0.0 & 6.260 &   -4.26470 &   -3.70710 &   -1.14680 &   -0.42180 &    0.09820 &    0.33490 &    0.81310 &    0.50650 &    0.52480 \\ 
4056.19 & 2.00 &  0.5 & 6.276 &   -4.61120 &   -4.01780 &   -1.21860 &   -0.47660 &    0.10630 &    0.36120 &    0.83980 &    0.59110 &    0.65440 \\
\end{longtable}
\end{landscape}
}

\longtab{
\scriptsize
\begin{landscape}
\label{simus3}
\begin{longtable}{c c c | c | c c c c c c c c c c c c c c c}
\caption{RHD simulations' stellar parameters (first three columns), bolometric Magnitude (M$_{\rm{bol}}$), and bolometric correction ($BC$) for the HST-WFC3 in VEGA system (Table~\ref{tablefilters}).  }\\
\hline\hline
$T_{\rm{eff}}$ & log$g$ & [Fe/H] & $M_{\rm{bol}}$ & $BC_{\rm{F218W}}$ & $BC_{\rm{F225W}}$ & $BC_{\rm{F275W}}$ & $BC_{\rm{F336W}}$ & $BC_{\rm{F350LP}}$ & $BC_{\rm{F390M}}$ & $BC_{\rm{F390W}}$ & $BC_{\rm{F475W}}$ & $BC_{\rm{F547M}}$ & $BC_{\rm{F555W}}$ & $BC_{\rm{F606W}}$ & $BC_{\rm{F625W}}$ & $BC_{\rm{F775W}}$ & $BC_{\rm{F814W}}$ & $BC_{\rm{F850LP}}$  \\
\hline
4014.99 & 1.50 &  0.0 & 6.320 &   -8.17260 &   -8.21110 &   -7.35370 &   -4.08830 &   -0.58120 &   -3.90800 &   -3.09130 &   -1.64640 &   -0.90720 &   -1.02750 &   -0.53660 &   -0.26000 &    0.43210 &    0.54620 &    0.85600 \\ 
4042.38 & 1.50 & -1.0 & 6.291 &   -7.87280 &   -7.75180 &   -6.50360 &   -3.39310 &   -0.56110 &   -3.35290 &   -2.74100 &   -1.56500 &   -0.87000 &   -0.98930 &   -0.52060 &   -0.25350 &    0.43310 &    0.54400 &    0.84560 \\ 
4021.90 & 1.50 & -2.0 & 6.313 &   -7.75910 &   -7.54350 &   -6.44760 &   -3.65980 &   -0.57650 &   -3.30190 &   -2.86240 &   -1.64070 &   -0.90510 &   -1.03570 &   -0.54250 &   -0.27040 &    0.44110 &    0.55510 &    0.86230 \\ 
3951.92 & 1.50 & -3.0 & 6.389 &   -8.12020 &   -7.88190 &   -7.07720 &   -4.48040 &   -0.63480 &   -3.78300 &   -3.36510 &   -1.90400 &   -1.05660 &   -1.20320 &   -0.63410 &   -0.33660 &    0.45520 &    0.57880 &    0.90650 \\ 
4070.69 & 2.00 &  0.0 & 6.260 &   -8.01960 &   -8.01340 &   -6.94350 &   -3.75960 &   -0.54480 &   -3.63140 &   -2.89600 &   -1.56440 &   -0.85310 &   -0.96900 &   -0.49470 &   -0.22220 &    0.45060 &    0.56060 &    0.85950 \\ 
4056.19 & 2.00 &  0.5 & 6.276 &   -8.15980 &   -8.18280 &   -7.31620 &   -4.23390 &   -0.57620 &   -3.94190 &   -3.10020 &   -1.64450 &   -0.91540 &   -1.03720 &   -0.54940 &   -0.27420 &    0.45290 &    0.57010 &    0.88610 \\
\end{longtable}
\end{landscape}
}

\longtab{
\scriptsize
\begin{landscape}
\label{simus4}
\begin{longtable}{c c c | c | c c c c c c c c c c c c c c c}
\caption{RHD simulations' stellar parameters (first three columns), bolometric Magnitude (M$_{\rm{bol}}$), and bolometric correction ($BC$) for the HST-WFC3 in ST system (Table~\ref{tablefilters}).  }\\
\hline\hline
$T_{\rm{eff}}$ & log$g$ & [Fe/H] & $M_{\rm{bol}}$ & $BC_{\rm{F218W}}$ & $BC_{\rm{F225W}}$ & $BC_{\rm{F275W}}$ & $BC_{\rm{F336W}}$ & $BC_{\rm{F350LP}}$ & $BC_{\rm{F390M}}$ & $BC_{\rm{F390W}}$ & $BC_{\rm{F475W}}$ & $BC_{\rm{F547M}}$ & $BC_{\rm{F555W}}$ & $BC_{\rm{F606W}}$ & $BC_{\rm{F625W}}$ & $BC_{\rm{F775W}}$ & $BC_{\rm{F814W}}$ & $BC_{\rm{F850LP}}$  \\
\hline
4014.99 & 1.50 &  0.0 & 6.320 &   -7.91180 &   -8.05600 &   -7.32530 &   -4.21030 &   -0.88440 &   -3.26900 &   -2.58400 &   -1.25150 &   -0.89580 &   -0.93570 &   -0.77710 &   -0.69240 &   -0.67450 &   -0.70340 &   -0.78440 \\ 
4042.38 & 1.50 & -1.0 & 6.291 &   -7.61200 &   -7.59670 &   -6.47520 &   -3.51520 &   -0.86420 &   -2.71400 &   -2.23380 &   -1.17010 &   -0.85850 &   -0.89760 &   -0.76120 &   -0.68580 &   -0.67350 &   -0.70570 &   -0.79480 \\ 
4021.90 & 1.50 & -2.0 & 6.313 &   -7.49830 &   -7.38840 &   -6.41920 &   -3.78190 &   -0.87970 &   -2.66300 &   -2.35520 &   -1.24580 &   -0.89370 &   -0.94400 &   -0.78310 &   -0.70280 &   -0.66550 &   -0.69460 &   -0.77810 \\ 
3951.92 & 1.50 & -3.0 & 6.389 &   -7.85940 &   -7.72680 &   -7.04880 &   -4.60250 &   -0.93800 &   -3.14400 &   -2.85790 &   -1.50910 &   -1.04510 &   -1.11150 &   -0.87470 &   -0.76900 &   -0.65140 &   -0.67080 &   -0.73380 \\ 
4070.69 & 2.00 &  0.0 & 6.260 &   -7.75870 &   -7.85830 &   -6.91510 &   -3.88170 &   -0.84790 &   -2.99240 &   -2.38880 &   -1.16940 &   -0.84160 &   -0.87730 &   -0.73520 &   -0.65460 &   -0.65590 &   -0.68900 &   -0.78080 \\ 
4056.19 & 2.00 &  0.5 & 6.276 &   -7.89890 &   -8.02770 &   -7.28780 &   -4.35600 &   -0.87930 &   -3.30300 &   -2.59300 &   -1.24950 &   -0.90400 &   -0.94540 &   -0.79000 &   -0.70650 &   -0.65360 &   -0.67950 &   -0.75430 \\
\end{longtable}
\end{landscape}
}

\longtab{
\scriptsize
\begin{landscape}
\label{simus5}
\begin{longtable}{c c c | c | c c c c c c c c c c c c c c c}
\caption{RHD simulations' stellar parameters (first three columns), bolometric Magnitude (M$_{\rm{bol}}$), and bolometric correction ($BC$) for the HST-WFC3 in AB system (Table~\ref{tablefilters}).  }\\
\hline\hline
$T_{\rm{eff}}$ & log$g$ & [Fe/H] & $M_{\rm{bol}}$ & $BC_{\rm{F218W}}$ & $BC_{\rm{F225W}}$ & $BC_{\rm{F275W}}$ & $BC_{\rm{F336W}}$ & $BC_{\rm{F350LP}}$ & $BC_{\rm{F390M}}$ & $BC_{\rm{F390W}}$ & $BC_{\rm{F475W}}$ & $BC_{\rm{F547M}}$ & $BC_{\rm{F555W}}$ & $BC_{\rm{F606W}}$ & $BC_{\rm{F625W}}$ & $BC_{\rm{F775W}}$ & $BC_{\rm{F814W}}$ & $BC_{\rm{F850LP}}$  \\
\hline
4014.99 & 1.50 &  0.0 & 6.320 &   -9.86380 &   -9.87180 &   -8.85240 &   -5.27410 &   -0.73790 &   -4.00730 &   -3.30720 &   -1.54930 &   -0.90690 &   -1.00310 &   -0.61960 &   -0.40810 &    0.05140 &    0.12800 &    0.33510 \\ 
4042.38 & 1.50 & -1.0 & 6.291 &   -9.56400 &   -9.41250 &   -8.00220 &   -4.57890 &   -0.71780 &   -3.45220 &   -2.95700 &   -1.46790 &   -0.86960 &   -0.96490 &   -0.60360 &   -0.40150 &    0.05230 &    0.12570 &    0.32480 \\ 
4021.90 & 1.50 & -2.0 & 6.313 &   -9.45030 &   -9.20420 &   -7.94620 &   -4.84560 &   -0.73320 &   -3.40120 &   -3.07830 &   -1.54360 &   -0.90480 &   -1.01130 &   -0.62550 &   -0.41840 &    0.06040 &    0.13680 &    0.34150 \\ 
3951.92 & 1.50 & -3.0 & 6.389 &   -9.81140 &   -9.54260 &   -8.57580 &   -5.66620 &   -0.79160 &   -3.88230 &   -3.58110 &   -1.80690 &   -1.05620 &   -1.17880 &   -0.71710 &   -0.48460 &    0.07440 &    0.16060 &    0.38570 \\ 
4070.69 & 2.00 &  0.0 & 6.260 &   -9.71080 &   -9.67410 &   -8.44210 &   -4.94540 &   -0.70150 &   -3.73070 &   -3.11200 &   -1.46720 &   -0.85270 &   -0.94470 &   -0.57770 &   -0.37030 &    0.06990 &    0.14230 &    0.33870 \\ 
4056.19 & 2.00 &  0.5 & 6.276 &   -9.85100 &   -9.84350 &   -8.81480 &   -5.41970 &   -0.73290 &   -4.04120 &   -3.31610 &   -1.54730 &   -0.91500 &   -1.01280 &   -0.63240 &   -0.42220 &    0.07220 &    0.15190 &    0.36520 \\
\end{longtable}
\end{landscape}
}

\begin{acknowledgements}
L.C. gratefully acknowledges support from the Australian Research Council (grants DP150100250, FT160100402). This work was granted access to the HPC resources of Observatoire de la C\^ote d'Azur - M\'esocentre SIGAMM.
\end{acknowledgements}

   \bibliographystyle{aa}
\bibliography{biblio.bib}

\begin{appendix} 
\section{One-dimensional versus three-dimensional bolometric corrections }

\begin{figure*}
   \centering
    \begin{tabular}{c}
       \includegraphics[width=1.0\hsize]{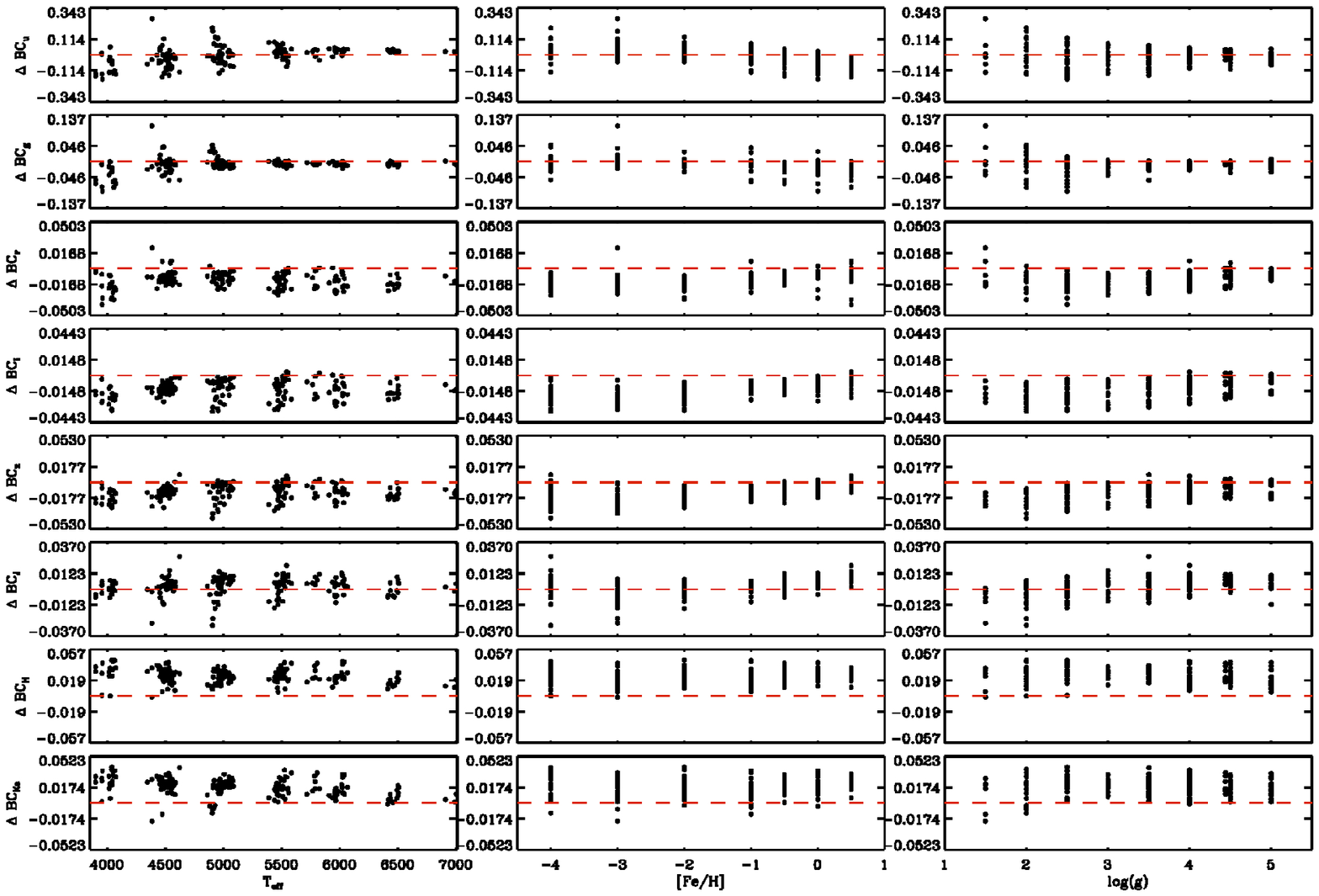} \\
    \end{tabular}
      \caption{Bolometric correction (BC) differences computed for the photometric filters Johnson-Cousins and 2MASS with 3D simulations (Table~4) and the corresponding 1D hydrostatic models with microturbulence = 1km/s: $\Delta BC= BC_{\rm{1D}}-BC_{\rm{3D}}$. The red dashed line indicates $\Delta BC=0.0$.}
        \label{BCcolors1}
           \end{figure*}  

\begin{figure*}
   \centering
    \begin{tabular}{c}
       \includegraphics[width=1.0\hsize]{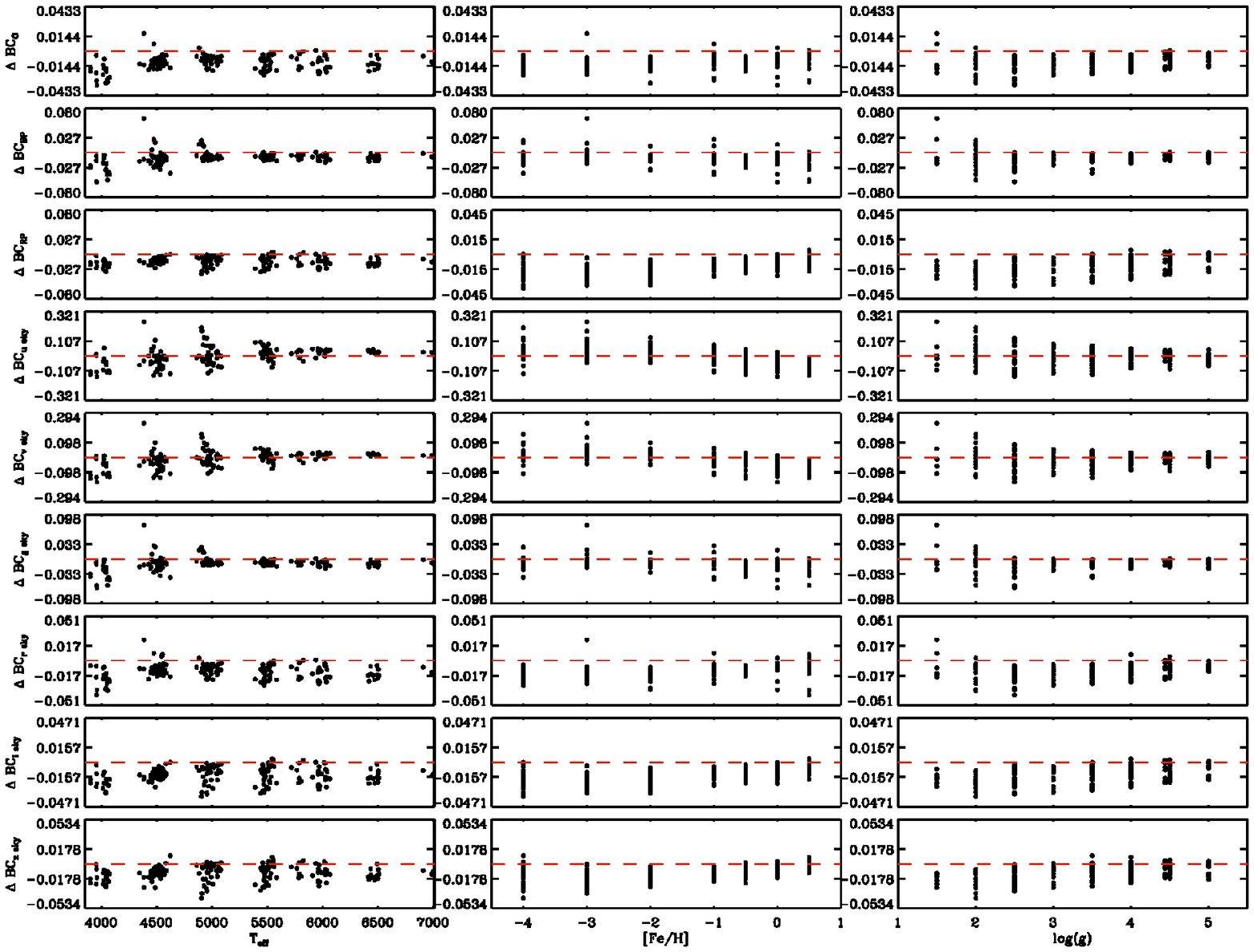} \\
    \end{tabular}
      \caption{Bolometric correction differences for the SkyMapper and Gaia photometric filters (Table~\ref{tablefilters}). The notation is the same as in Fig.~\ref{BCcolors1}. }
        \label{BCcolors2}
           \end{figure*}  

\begin{figure*}
   \centering
    \begin{tabular}{c}
       \includegraphics[width=1.0\hsize]{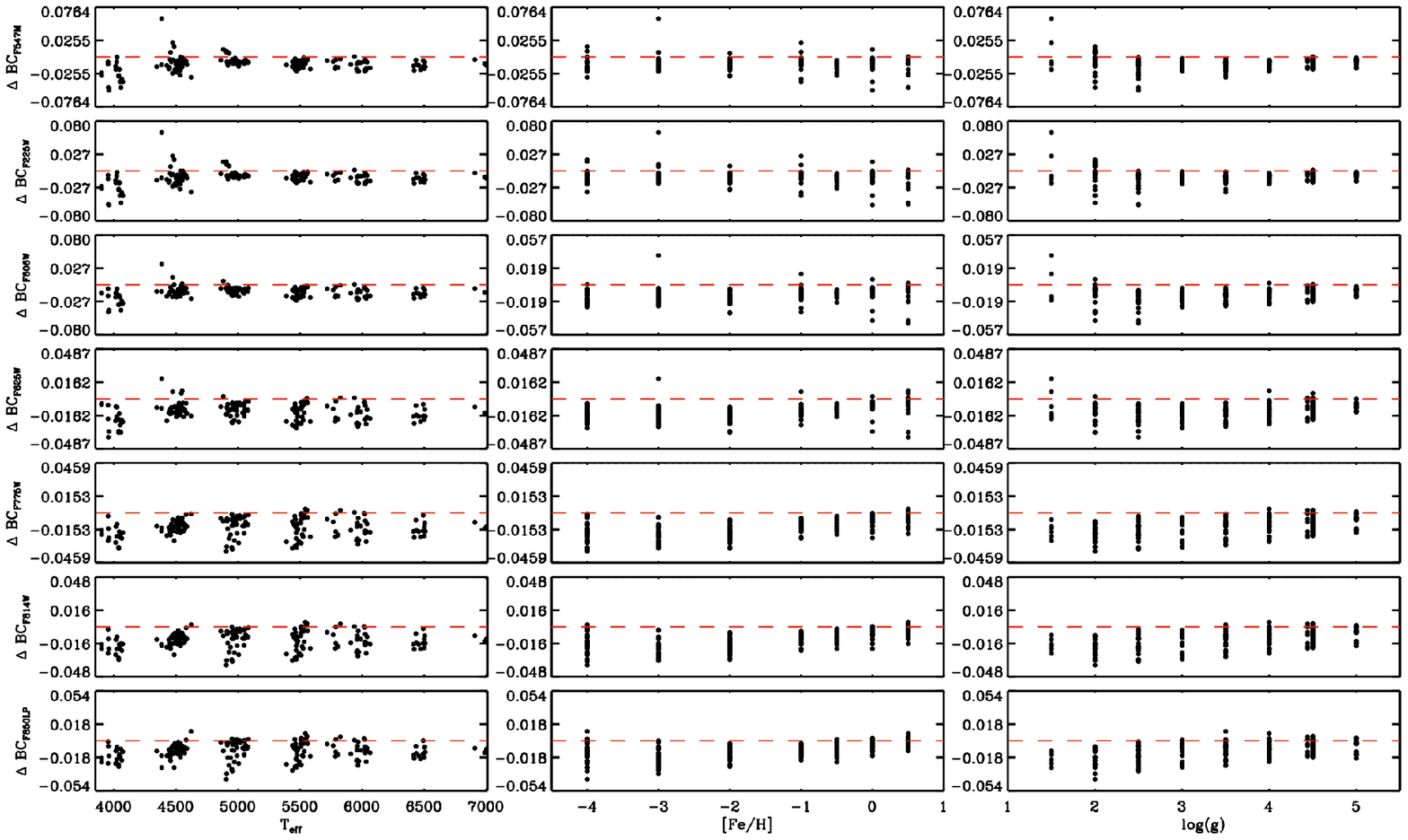} \\
    \end{tabular}
      \caption{Bolometric correction differences for the HST-WFC3 provided in the VEGA system (for the ST and AB systems the differences are similar);  first set of filters  (Table~\ref{tablefilters}). The notation is the same as in Fig.~\ref{BCcolors1}.}
           \label{BCcolors3}
           \end{figure*}  
    
    \begin{figure*}
   \centering
    \begin{tabular}{c}
       \includegraphics[width=1.0\hsize]{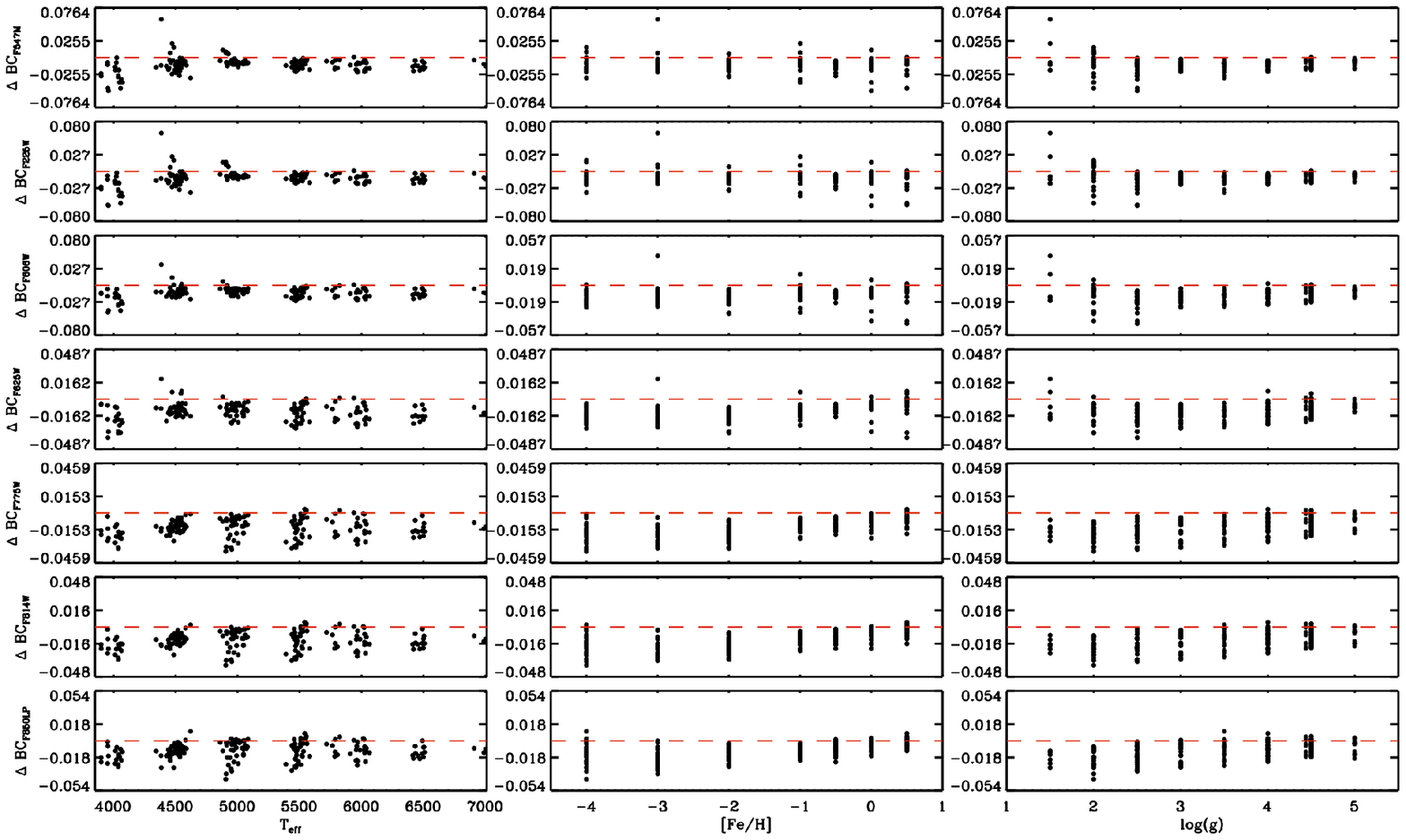} \\
    \end{tabular}
      \caption{Bolometric correction differences for the HST-WFC3 provided in the VEGA system (for ST and AB systems the differences are similar);  second set of filters (Table~\ref{tablefilters}). The notation is the same as in Fig.~\ref{BCcolors1}.}
           \label{BCcolors4}
           \end{figure*}

\end{appendix}

\end{document}